# Response-Based Frequency Stability Assessment under Multi-Scale Disturbances in High-Renewable Power Systems

Jinhui Chen, *Student Member, IEEE,* Huadong Sun, *Senior Member, IEEE, Fellow, CSEE,* Ping Wu, Baocai Wang, and Bing Zhao, *Senior Member, IEEE*

***Abstract*—In high-renewable power systems, active-power disturbances are becoming larger and more diverse in time scale, which complicates frequency stability assessment under unanticipated events. This paper proposes a response-based frequency stability assessment method that infers disturbance power from measured generator electrical responses and enables a unified treatment of multi-scale disturbances. Using the inferred disturbance power and its rate of change, unanticipated events are classified into short-term and permanent disturbances, where permanent disturbances are further categorized as step, second-level slope, and minute-level slope disturbances. For each disturbance class, analytical frequency-response criteria are derived, including an inertial energy-accumulation criterion for short-term disturbances, an improved system frequency response (SFR) model for step and second-level slope disturbances, and a post-exhaustion rotor-motion model for minute-level slope disturbances. The resulting criteria provide quantitative stability margins, including critical disturbance limits and time-to-limit prediction. Case studies on the CSEE-FS frequency stability benchmark system, together with comparisons against a conventional SFR-based assessment method, validate the correctness and effectiveness of the proposed approach.**

## I. Introduction

With the large-scale integration of renewable generation and HVDC transmission, synchronous inertia and conventional primary frequency regulation capability are steadily declining, making the system frequency response more vulnerable to disturbances [1], [2]. Meanwhile, active-power disturbances are becoming increasingly diverse in duration and temporal evolution, posing new challenges to frequency stability [3], [4].

For conventional step disturbances with known magnitude, existing frequency-stability studies have provided a solid analytical basis. The classical SFR model in [5] gives low-order representations of inertial response and primary frequency regulation, and has been widely used for frequency nadir prediction and fast security screening. Building on this foundation, subsequent studies have refined low-order frequency-response representations for improved nadir prediction and have extended the SFR framework to high-renewable systems [6], [7]. Closed-form approximations of the maximum frequency deviation have also been developed based on linearized primary regulation, providing efficient tools for planning and security assessment [8], [9]. The same modeling framework has also been extended to capability-oriented assessment, including the evaluation of frequency-support capability for diverse renewable units and explicit consideration of thermal-unit dynamic limits [10], [11]. These studies show that low-order frequency-response models are extensible and can provide analytical tools for frequency-security assessment under changing generation structures.

However, practical disturbance profiles have become more diverse, and frequency security can be dominated by short-term disturbances and time-varying slope disturbances. Short-term disturbances are often triggered by fast power-transfer reductions during LCC-HVDC commutation failures, or by temporary wind-power suppression during low-voltage ride-through (LVRT), especially when post-fault active-power recovery is intentionally delayed by turbine controls, producing a short but severe generation shortfall [12], [13]. Time-varying slope disturbances can be interpreted as a sequence of successive losses and recoveries that yields a nonzero net slope, ranging from second-level slopes caused by cascading trips to minute-level slopes driven by sustained load growth. A representative second-level case is the 9 August 2019 Great Britain event, where multiple sources tripped in close succession and the accelerating frequency decline triggered low-frequency demand disconnection [14]. In contrast, the September 2021 Northeast China event reflects a slower yet persistent supply–demand gap, where frequency declined to below 49.8 Hz and emergency load shedding had to be implemented after available regulation measures were exhausted [15].

In parallel, online measurement-based applications have provided new opportunities for frequency-stability assessment. Inertia estimation has progressed toward response-driven

This work was supported by the Smart Grid-National Science and Technology Major Project (Demonstration Project on Construction and Security Control of Renewable Energy-Dominant, UHV-Intensive Regional Sending-End Systems, 2025ZD0806900). J. H. Chen, H. D. Sun (corresponding author, email: sunhd@epri.sgcc.com.cn), P. Wu, B. C. Wang and B. Zhao are with the State Key Laboratory of Power Grid Safety and Energy Conservation, China Electric Power Research Institute, Beijing 100192, China.

approaches that infer system-level inertia from measured frequency trajectories under events or ambient conditions [16]–[19]. Recent studies have also addressed practical cyber and communication constraints in frequency-related applications, including resilience enhancement against cyber attacks and delay compensation for reporting and control signals in frequency regulation [20]–[23]. With the development of synchrophasor and wide-area measurement technologies, electrical quantities in power systems can be measured with improved accuracy and time synchronization. PMU-based studies further show that synchronized measurements can support online applications such as oscillation source location and voltage-stability assessment, while measurement-error characterization and compensation can improve the reliability of PMU-based monitoring applications [24]–[26]. These advances highlight the need for response-based, quantitative tools that can directly link disturbance characteristics extracted from measured responses with frequency-stability margins in high-renewable power systems.

The representative frequency-stability studies reviewed above are summarized in Table I. Existing studies have provided effective tools for frequency-response prediction, frequency-support assessment, inertia-related assessment, and resilient regulation. However, most of them either assume the disturbance form in advance or use disturbances only as estimation and control inputs. A response-based quantitative framework that can identify unanticipated disturbance types and compute class-dependent frequency-stability margins in real time remains lacking, especially for time-varying slope disturbances. This paper proposes a response-based frequency stability assessment method for multi-scale unanticipated disturbances. The main contributions are as follows.

1) A unified response-based disturbance classification scheme is developed to identify short-term, step, second-level slope, and minute-level slope disturbances using the disturbance power and its rate of change inferred from generator electrical responses.

2) Analytical criteria are derived for each disturbance class. For step disturbances, the maximum tolerable disturbance power is obtained under steady-state and transient frequency-deviation constraints, and a safety-margin index is defined. For slope disturbances, an improved SFR model and a post-exhaustion rotor-motion model are used to predict the time to frequency-limit violation.

3) The method is validated on the CSEE-FS benchmark system under LVRT, step-load, and slope scenarios [27]. Comparisons against conventional SFR-based benchmarks are included. For minute-level slope disturbances, comparison with the conventional SFR-based method is performed using the same slope disturbance input rather than an equivalent step approximation, thereby avoiding the additional bias introduced by step substitution and highlighting the error caused by neglecting the transition after primary-frequency-regulation exhaustion.

TABLE I
COMPARISON OF RELATED WORKS AND DISTURBANCE ASSUMPTIONS

| Refs. | Topic | Method or model | Assumed disturbance form | Timescale | Comments |
|---|---|---|---|---|---|
| [5]–[9] | Low-order frequency response and nadir prediction | SFR models and analytical nadir calculation | Known generation or load imbalance, usually represented as a permanent step disturbance | Seconds to tens of seconds | Effective for known step disturbances, but not for unanticipated short-term or time-varying slope disturbances |
| [10], [11] | Frequency-support capability and realistic limits | Renewable support assessment and thermal-unit dynamic limits | Step-like contingencies with support or regulation limits | Seconds to minutes | Considers realistic support limits, but lacks response-based disturbance-type discrimination |
| [12], [13] | Short-term disturbance and recovery mechanisms | HVDC commutation-failure and LVRT recovery analysis | Short-term disturbance followed by post-fault recovery | Sub-second to seconds | Shows short-term frequency-security risks, but remains mechanism- or case-oriented |
| [14], [15] | Time-varying slope disturbance evidence | Post-event analysis and resource-adequacy reports | Successive losses, delayed recovery, or persistent supply-demand gap that may form a time-varying slope disturbance | Seconds to minutes and longer | Shows evolving disturbances in real systems, but lacks a reusable response-based assessment method |
| [16]–[19] | System inertia estimation | Event-based, ambient-data-based, and maximum-deviation-based estimation | Frequency events or ambient fluctuations used only as estimation inputs | First seconds to continuous operation | Estimates inertia, not disturbance form, disturbance type, or real-time frequency-stability margin |
| [20]–[23] | Cyber threats and communication latency | Resilient load frequency control, attack mitigation, and delay compensation | Step or random load disturbances with attacks or delays | Seconds to minutes | Improves frequency-regulation resilience, but not disturbance-type-based stability assessment |
| [24]–[26] | PMU-based online monitoring and measurement-error modeling | Oscillation source location, systematic error compensation, and PMU error characterization | Measured events used as inputs, without disturbance-type classification | Seconds to continuous monitoring | Supports reliable response-based monitoring, but not disturbance-type-based frequency-stability assessment |
| Proposed | Multi-scale unanticipated disturbance assessment | Response-based disturbance inference and analytical criteria | Short-term, step, second-level slope, and minute-level slope disturbances inferred from responses | Sub-second to minutes | Identifies disturbance type online and quantifies frequency-stability margin or time-to-limit violation |

## II. Disturbance Type Identification

### *A. Disturbance Scenario Classification*

Unanticipated disturbances are first classified into short-term disturbances and permanent disturbances. A short-term disturbance is characterized by a large disturbance power that rises rapidly and then returns close to the pre-disturbance level within a short time interval. Permanent disturbances are further divided into three categories.

1) Step disturbance. The disturbance power experiences a sudden increase and then remains approximately constant. Its severity is quantified by the disturbance power magnitude.

2) Second-level slope disturbance. The disturbance power increases approximately linearly with a relatively large slope, and the transient frequency-deviation limit can be reached within several seconds. Its severity is quantified by the disturbance-power slope.

3) Minute-level slope disturbance. The disturbance power increases approximately linearly but with a small slope over a long duration, and the time scale to reach the frequency-deviation limit is on the order of minutes. Its severity is quantified by the disturbance-power slope.

Fig. 1 illustrates the relationship between the time scale of disturbance evolution and the time scale of system frequency response. According to the classical understanding of frequency dynamics, the system frequency response after a power imbalance can be divided into several successive stages, including the initial inertial response, primary frequency regulation, and recovery processes [28].

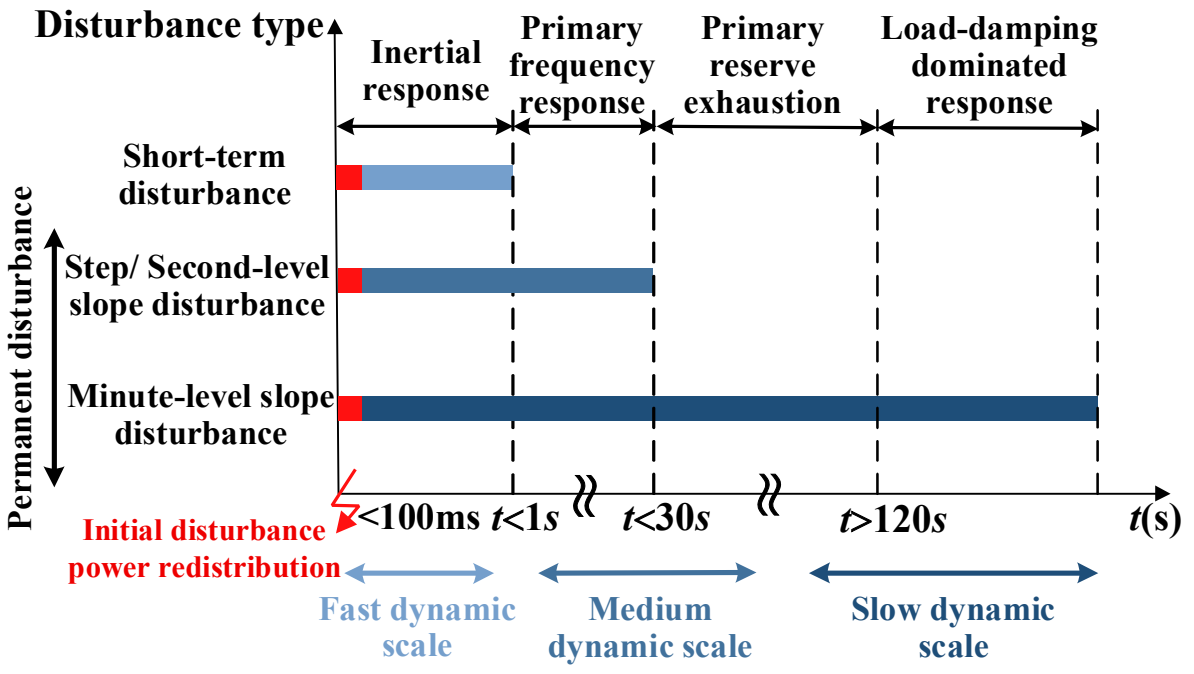

Fig. 1. Schematic relationship between multi-scale disturbance types and multi-time-scale frequency response.

As shown in Fig. 1, short-term disturbances evolve within a very short interval, typically on the order of several hundred milliseconds. Their frequency impact is mainly governed by the inertial response. Step disturbances and second-level slope disturbances develop over a longer transient interval, usually from the inertial response stage to the primary frequency-regulation stage. Therefore, their frequency-stability impact depends on both the initial inertial frequency decline and the subsequent primary frequency support. Minute-level slope disturbances are different from the above transient disturbances. Their disturbance power increases slowly over a much longer duration, and the early frequency deviation may be small. However, the accumulated imbalance can gradually consume the available primary frequency-regulation capability. Once the regulation capability is exhausted, the frequency will decline with a steeper slope until additional control actions are taken.

This classification clarifies the fundamental difference among disturbances of different scales. The difference does not only lie in the disturbance magnitude, but also in the rate and duration of disturbance-power variation. Different disturbance scales interact with different stages of frequency response, which leads to different dominant stability-evaluation quantities.

### *B. Identification of Disturbance Type*

The disturbance power is redistributed among generators in the early stage according to the network electrical distance and the instantaneous rotor-angle differences. This redistribution is reflected in the abrupt changes of generator electromagnetic power, which provides the response information required for disturbance-type identification. Therefore, the disturbance type can be identified from the estimated disturbance power and its time variation.

For step disturbances and second-level slope disturbances, the disturbance power grows rapidly during the early stage. This feature enables response-based identification using the estimated disturbance power and its time variation. Events are further classified as short-term disturbances if the estimated disturbance power subsequently decreases rapidly and returns close to its pre-disturbance value. For minute-level slope disturbances, the disturbance power does not exceed the predefined power threshold during the early stage, and the frequency therefore drifts slowly until the primary frequency-regulation capability is exhausted and the frequency deviation exceeds a specified threshold.

The resulting thresholding logic and its role in disturbance-type identification are illustrated in Fig. 2. If the estimated disturbance power exceeds $\Delta P_{\mathrm{sh}}$ within the early-stage disturbance-identification window $T$, the event is classified as either a step disturbance or a second-level slope disturbance. The two cases are then distinguished by whether the estimated disturbance power exhibits a clear slope trend over $T$: a slope behavior indicates a second-level slope disturbance, whereas an approximately constant value indicates a step disturbance. The event is finally classified as a short-term disturbance if the estimated disturbance power drops rapidly after fault clearing and returns close to its pre-disturbance level.

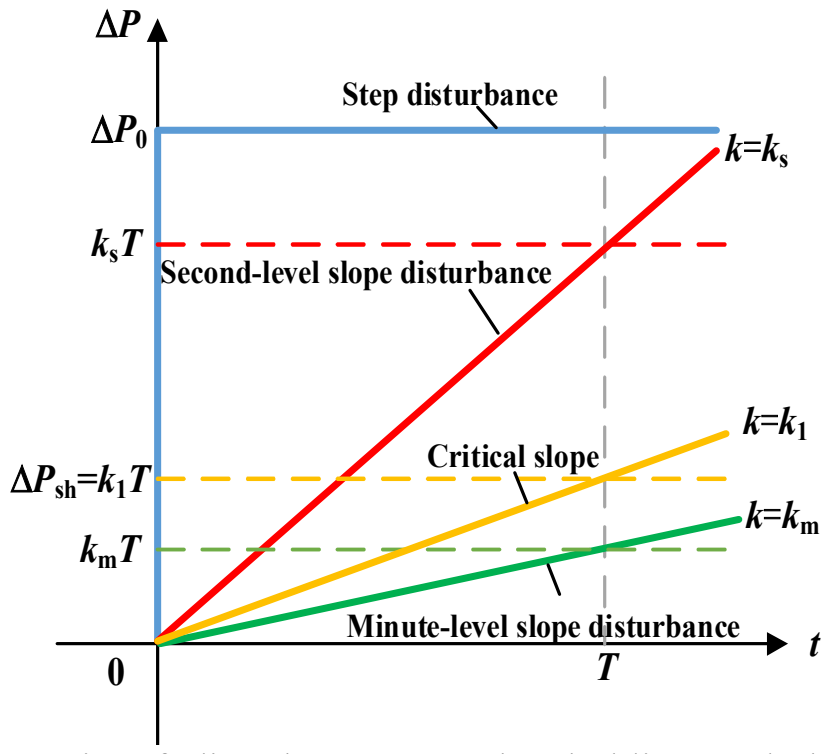

Fig. 2. Schematic of disturbance-type thresholding and the setting of the disturbance power threshold $\Delta P_{\mathrm{sh}}$.

### C. Threshold Determination

To accurately distinguish the four types of disturbances, the power threshold must be set according to the system operating condition. The power threshold is introduced to avoid misclassifying minute-level slope disturbances as step or second-level slope disturbances. Specifically, it is selected as the maximum disturbance power that a minute-level slope disturbance can produce within $T$.

The characteristic slopes of second-level and minute-level slope disturbances are summarized in Table II. Let $k_1$ denote the critical disturbance-power slope separating these two classes. For a minute-level slope disturbance, the maximum disturbance power accumulated within the early-stage window is $k_1T$. Accordingly, the power threshold is set as $\Delta P_{sh} = k_1T$.

TABLE II
RANGES OF DISTURBANCE POWER SLOPE FOR SECOND-LEVEL AND MINUTE-LEVEL SLOPE DISTURBANCES

| Disturbance type | Disturbance power slope $k$ (MW/s) |
|---|---|
| Second-level slope disturbance | $k > k_1$ |
| Minute-level slope disturbance | $0 < k \leq k_1$ |

The frequency-deviation threshold $\Delta f_{sh}$ is defined as the frequency deviation at which the primary frequency-regulation capacity of the generators is fully exhausted.

Assume that the primary frequency-regulation capacity of synchronous generators accounts for a proportion $m$ of their total rated capacity, and that renewable resources provide a proportion $n$. When the generator primary frequency-regulation capacity is fully utilized, the corresponding frequency deviations can be expressed as (1) and (2):

$$\Delta f_{\mathrm{G}} = \frac{mP_{\mathrm{GN}} + nP_{\mathrm{NEW}}}{(K_{\mathrm{G}} + K_{\mathrm{w}})\,S_{\mathrm{N}}} f_{\mathrm{N}} \tag{1}$$

$$\Delta f_{\mathrm{sh}} = \Delta f_{\mathrm{G}} + f_{\mathrm{d}} \tag{2}$$

where $\Delta f_G$ is the frequency deviation at which the available generator primary frequency regulation is fully deployed, $K_G$ and $K_w$ are the unit regulation gains of the generators and the renewable resources, respectively, $P_{GN}$ and $P_{NEW}$ are the rated capacities of the generators and renewable resources, respectively, $S_N$ is the system power base taken as the sum of $P_{GN}$ and $P_{NEW}$, $f_d$ is the frequency deviation at which primary frequency control is activated, and $\Delta f_{sh}$ denotes the total frequency deviation when the generator primary frequency-regulation capacity is exhausted.

When $|\Delta f| \geq \Delta f_{sh}$ and the estimated disturbance power does not reach the power threshold, the disturbance is identified as a minute-level slope disturbance. In this case, the disturbance-power slope is determined from regulation limits together with the power-balance condition at the threshold instant.

As an example, consider the parameter set: $K_G = 20$, $K_w = 10$, $K_L = 1$, $H = 4$, $P_{GN} = 1000$ MW, $P_{NEW} = 1000$ MW, $f_d = 0.033$ Hz, $m = 0.06$, $n = 0.1$. Suppose that the time required for a minute-level slope disturbance to reach the frequency-activation threshold exceeds 30 s. The corresponding critical slope is then $k_1 = 3$ MW/s, meaning that the disturbance power increases at a rate of 0.15% of the system capacity per second. If $T = 0.5$ s, the resulting power threshold is $\Delta P_{sh} = 1.5$ MW, and the frequency deviation threshold is $\Delta f_{sh} = 0.3$ Hz.

## III. QUANTIFICATION OF DISTURBANCE INTENSITY

### A. Short-Term and Step Disturbances

For short-term disturbances and step disturbances, the disturbance intensity is quantified by the dominant magnitude of the inferred disturbance power within the early-stage identification window $T$, which can be written as:

$$\Delta P_0 = \max_{t\in[t_0,t_0+T]} \left|\Delta P_{\mathrm{d}}(t)\right|, \quad \Delta P_{\mathrm{d}}(t) = \sum_{i=1}^{n} \Delta P_{\mathrm{e}i}(t) \tag{3}$$

where $\Delta P_0$ is the disturbance intensity, $\Delta P_d(t)$ is the inferred disturbance power, $\Delta P_{ei}(t)$ is the electromagnetic-power increment of generator $i$, $t_0$ is the disturbance inception instant.

Based on the network power-injection relationship, the electromagnetic power of generator $i$ can be expressed as:

$$P_{\mathrm{e}i}(t) = \sum_{\substack{j=1 \\ j\neq i,k}}^{n} E_i E_j B_{ij} \sin\delta_{ij} + U_{\mathrm{p}} E_i B_{i\mathrm{p}} \sin\delta_{i\mathrm{p}} \tag{4}$$

where $E_i$ and $E_j$ are the internal voltages of generators $i$ and $j$, $U_p$ is the voltage magnitude at the disturbance point $p$, $\delta_{ij}$ is the power-angle difference between generators $i$ and $j$, $\delta_{ip}$ is the power-angle difference between generator $i$ and the disturbance point, and $B_{ij}$ and $B_{ip}$ are the corresponding network susceptance.

The electromagnetic-power increment of generator $i$ is obtained by direct response differencing:

$$\Delta P_{\mathrm{e}i}(t) = P_{\mathrm{e}i}(t) - P_{\mathrm{e}i}(t_0^{-}) \tag{5}$$

Since $B_{ij} = B_{ji}$, and $\delta_{ij} = -\delta_{ji}$, the inter-machine power-exchange variations cancel pairwise in the aggregate summation. Thus, the aggregate expression is simplified to the power variation associated with the disturbance location, eliminating the generator-to-generator oscillatory power-exchange terms. The aggregate disturbance power can be inferred as:

$$\begin{aligned}\Delta P_{\mathrm{d}}(t) = &\sum_{i=1}^{n} U_{\mathrm{d}}(t) E_i(t) B_{i\mathrm{p}}(t) \sin\delta_{i\mathrm{p}}(t) \\ &- \sum_{i=1}^{n} U_{\mathrm{p}}(t_0^{-}) E_i(t_0^{-}) B_{i\mathrm{p}}(t_0^{-}) \sin\delta_{i\mathrm{p}}(t_0^{-})\end{aligned} \tag{6}$$

Accordingly, the intensity of short-term and step disturbances can be obtained in a response-based manner directly from generator electrical responses.

### B. Second-Level Slope Disturbances

For second-level slope disturbances, the disturbance intensity is characterized by the disturbance-power slope, denoted by $k_s$. After $\Delta P_d(t)$ is obtained from (6), $k_s$ is estimated by least-squares fitting:

$$k_{\mathrm{s}} = \frac{\sum_{\ell=1}^{N} (t_\ell - \bar{t})\left[\Delta P_{\mathrm{d}}(t_\ell) - \overline{\Delta P_{\mathrm{d}}}\right]}{\sum_{\ell=1}^{N} (t_\ell - \bar{t})^2} \tag{7}$$

where $t_\ell$ is the sampling instant in the fitting window, $N$ is the number of samples, $\bar{t}$ is the mean value of $t_\ell$, and $\overline{\Delta P_d}$ is the mean value of the inferred disturbance power.

Hence, $k_s$ can be quantified from the estimated disturbance power trajectory obtained from measured generator electrical-power responses, providing a response-based index of disturbance intensity for second-level slope disturbances.

### *C. Minute-Level Slope Disturbances*

For minute-level slope disturbances, the disturbance intensity is quantified by the disturbance-power slope $k_m$. In this scenario, frequency evolves on a much slower time scale, and the disturbance power within the initial redistribution interval does not exceed the power threshold. The event is therefore identified via the exhaustion-related frequency-deviation threshold $\Delta f_{sh}$, as defined in Section II.

Let $t_1$ denote the instant when the frequency deviation first reaches $\Delta f_{sh}$. At this instant, the available primary frequency regulation of synchronous generators is assumed to be fully utilized, and the remaining imbalance is compensated primarily by load frequency damping. The damping contribution is expressed as:

$$\Delta P_L = K_L \Delta f_{sh} \quad (8)$$

where $K_L$ is the load damping coefficient of the system.

Enforcing the active-power balance at $t_1$ gives:

$$k_m = (K_L \Delta f_{sh} + mP_{GN} + nP_{NEW}) / t_1 \quad (9)$$

Equation (9) determines $k_m$ by linking the disturbance intensity explicitly to primary frequency-regulation limits and the selected exhaustion threshold. This provides a capability-consistent intensity measure for minute-level slope disturbances.

## IV. Frequency Stability Discrimination

### *A. Short-Term Disturbances*

For short-term disturbances, the response of primary frequency control is typically delayed, and the initial frequency dynamics are dominated by the inertial response of synchronous generators. The resulting energy accumulation process is illustrated in Fig. 3.

Let the time of maximum transient frequency deviation be $t_m$. The accumulated energies associated with the disturbance power deficit and the load frequency modulation at $t_m$ are given by (10) and (11):

$$\Delta S_P = (\Delta P_0 + \Delta P_0')(\Delta t + \Delta t') / 2 \quad (10)$$

$$\Delta S_L = K_L \Delta f_m P_{L0} (\Delta t + \Delta t') / 2 \quad (11)$$

where $\Delta P_0$ is the disturbance power magnitude during the fault plateau, $\Delta t$ is the fault duration , $\Delta t'$ is the time interval from fault clearing to $t_m$, and $\Delta P_0'$ is the residual disturbance power deficit at $t_m$. Since the post-fault disturbance power recovers approximately linearly at rate $K_R$, the residual deficit satisfies $\Delta P_0' = \Delta P_0 - K_R \Delta t'$, $P_{L0}$ is the pre-disturbance load power, $\Delta f_m$ is the maximum transient frequency deviation.

The minimum kinetic energy that must be provided by the system inertia to sustain the frequency excursion is:

$$W_G = \Delta S_P - \Delta S_L \quad (12)$$

where $W_G$ is the kinetic energy available from synchronous machines, given by $W_G = 2HP_{GN}$.

At $t_m$, the imbalance power is first re-balanced by the inertial response together with the load frequency modulation:

$$\Delta P_0 = K_R \Delta t' + K_L \Delta f_m P_{L0} \quad (13)$$

Combining (12) and (13), and denoting the transient frequency deviation limit by $\Delta f_{max}$, the critical disturbance power for a short-term disturbance is obtained as:

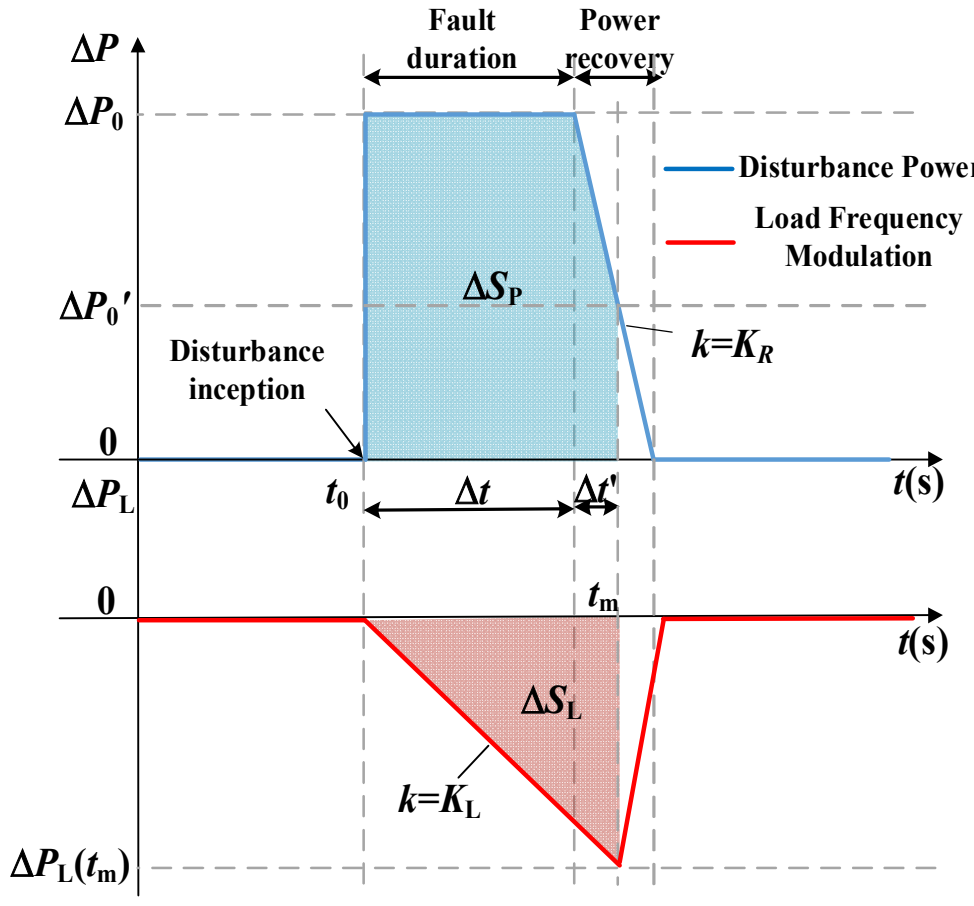

Fig. 3. Accumulated energy due to disturbance power and load-frequency modulation during a short-term disturbance.

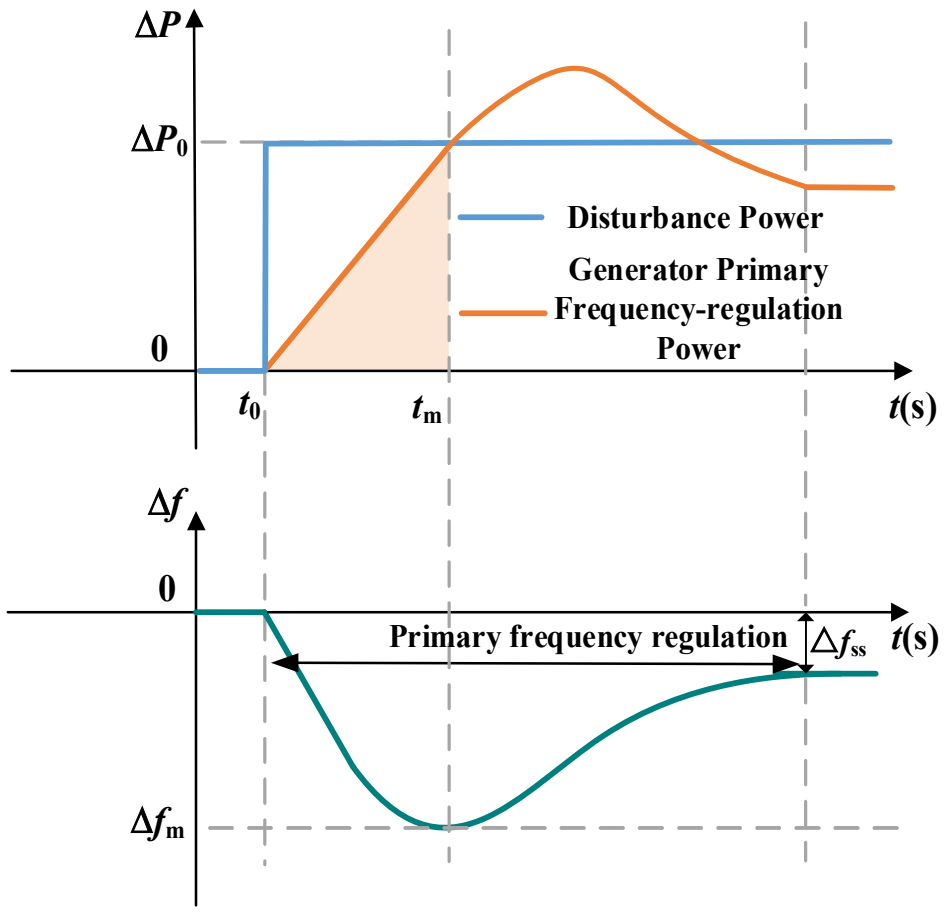

Fig. 4. Frequency response to a step disturbance.

$$\frac{\Delta P_{max}^2}{K_R} + \Delta P_{max} (\Delta t - \frac{K_L}{K_R} P_{L0} \Delta f_{max}) - 4H \Delta f_{max} P_{GN} = 0 \quad (14)$$

Equation (14) provides a quantitative criterion to assess whether a short-term disturbance will cause the transient frequency deviation to exceed the admissible limit.

### *B. Step Disturbances*

#### *1) Critical Disturbance Power Under Steady-State Frequency Deviation Constraint*

The steady-state frequency deviation is determined by generator primary frequency regulation and load frequency sensitivity. When the system reaches a new steady state after a step disturbance, the active-power imbalance equals the sum of the regulation contributions from generators and loads, yielding:

$$\Delta P_{max1} = (K_L \Delta f_{ss} P_{L0} + K_G \Delta f_{ss} P_{GN} + K_w \Delta f_{ss} P_{NEW}) / f_N \quad (15)$$

where $\Delta f_{ss}$ is admissible steady-state frequency-deviation limit, $f_N$ is the nominal system frequency (50 Hz), and $\Delta P_{max1}$ denotes the critical disturbance power under the steady-state constraint.

Since generator primary regulation capability is limited by prime-mover and reserve constraints, once generator regulation saturates, any additional imbalance must be compensated by load frequency regulation. The corresponding limiting power-balance condition is:

$$\Delta P_{max2} = mP_{GN} + nP_{NEW} + K_L \Delta f_{ss} P_{L0} / f_N \quad (16)$$

The smaller value obtained from (15) and (16) is taken as the maximum tolerable disturbance power under the steady-state frequency-deviation constraint.

*2) Critical Disturbance Power Under Transient Frequency Deviation Constraint*

The transient frequency deviation is governed jointly by system inertia and frequency-regulation capability. The maximum deviation occurs at the first instant when generation and load re-establish power balance following the disturbance, as illustrated in Fig. 4.

The SFR model is widely used to predict post-disturbance frequency dynamics. Following [7], an equivalent single-machine representation that includes the participation of power-electronic units is shown in Fig. 5.

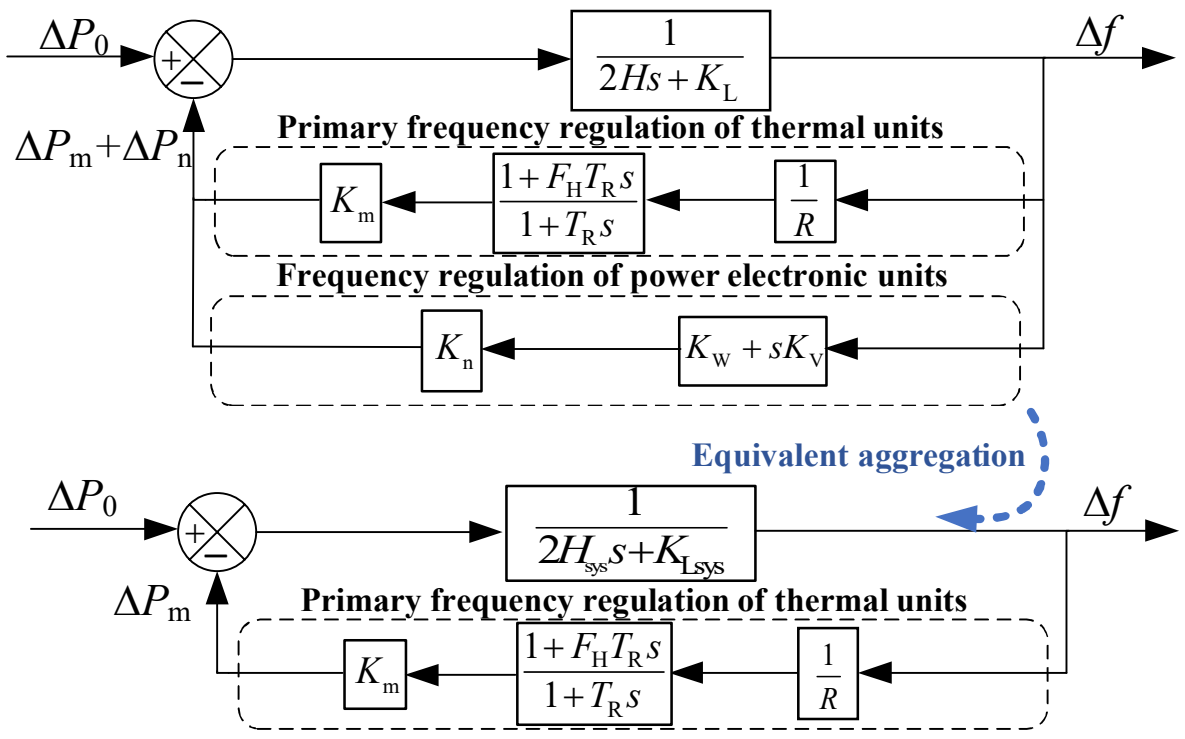

Fig. 5. SFR model with participation of power electronic units.

The time $t_m$ at which the maximum frequency deviation occurs and the corresponding maximum deviation are given by:

$$t_{\mathrm{m}} = \frac{1}{\omega_{\mathrm{r}}}\tan^{-1}\left(\frac{\omega_{\mathrm{r}}T_{\mathrm{R}}}{\zeta\omega_{\mathrm{n}}T_{\mathrm{R}}-1}\right) \tag{17}$$

$$\Delta f_{\mathrm{m}} = \frac{R\Delta P_0}{K_{\mathrm{Lsys}}R+1}\cdot\left[1+\sqrt{1-\zeta^2}\,\alpha e^{-\zeta\omega_{\mathrm{n}}t_{\mathrm{m}}}\right] \tag{18}$$

where $K_{\mathrm{Lsys}} = K_{\mathrm{L}} + K_{\mathrm{n}}K_{\mathrm{w}}$, $H_{\mathrm{sys}} = H + K_{\mathrm{n}}K_{\mathrm{v}}/2$, $K_{\mathrm{n}}$ is the proportion of renewable resources, and $K_{\mathrm{v}}$ is the virtual inertia coefficient. The remaining variables and detailed expressions follow [5].

Based on (17)–(18), the maximum disturbance power that the system can tolerate under the transient frequency-deviation constraint is obtained as:

$$\Delta P_{\max 3} = \frac{\Delta f_{\max}(K_{\mathrm{Lsys}}R+1)S_{\mathrm{N}}}{R(1+\sqrt{1-\zeta^2}\,\alpha e^{-\zeta\omega_{\mathrm{n}}t_{\mathrm{m}}})} \tag{19}$$

Comparing the steady-state and transient constraints, the overall critical disturbance power for a step disturbance is determined as:

$$\Delta P_{max} = min\{\Delta P_{\max 1}, \Delta P_{\max 2}, \Delta P_{\max 3}\} \tag{20}$$

To quantify the disturbance severity and visualize the remaining frequency stability margin, a safety-margin indicator is defined as:

$$\eta = \frac{\Delta P_{\max} - \Delta P_0}{\Delta P_{\max}} \tag{21}$$

When $0 < \eta < 1$, the frequency remains within limits, and a smaller $\eta$ indicates a smaller safety margin. When $\eta < 0$, the frequency violates the limit, and a more negative $\eta$ corresponds to a more severe event.

## *C. Second-Level Slope Disturbances*

Once a second-level slope disturbance is identified, the time required for the frequency deviation to reach the transient limit is typically short. Therefore, the SFR model can be used to compute the over-limit time under the transient frequency-deviation constraint. The induced frequency response is illustrated in Fig. 6.

Assuming a linear growth of disturbance power with disturbance inception taken as the time origin, the disturbance power can be written as $\Delta P_0 = k_s t$. The modified SFR model for a slope disturbance in the Laplace domain is:

$$\Delta f(s) = \frac{k_s T_1 T_2 R\omega_{\mathrm{n}}^2}{K_{\mathrm{Lsys}}R+1}\left(\frac{1+T_{\mathrm{R}}s}{s^2(1+T_1 s)(1+T_2 s)}\right) \tag{22}$$

where $T_1 = \dfrac{\zeta+\sqrt{\zeta^2-1}}{\omega_{\mathrm{n}}}$, $T_2 = \dfrac{\zeta-\sqrt{\zeta^2-1}}{\omega_{\mathrm{n}}}$.

Applying the inverse Laplace transform to (22) yields the time-domain frequency deviation:

$$\Delta f(t) = \frac{k_s R}{K_{\mathrm{Lsys}}R+1}\left(t+T_{\mathrm{R}}-T_1-T_2+k_1e^{-t/T_1}+k_2e^{-t/T_2}\right) \tag{23}$$

where $k_1 = \dfrac{T_1^2 - T_R T_1}{T_1 - T_2}$, $k_2 = \dfrac{T_2^2 - T_R T_2}{T_2 - T_1}$.

Substituting $\Delta f(t) = \Delta f_{\max}$ and $t = t_{\mathrm{m}}$ into (23) yields:

$$\Delta f_{\max}(K_{\mathrm{Lsys}}R+1)/k_s R = t_{\mathrm{m}} + k_1 e^{-t_{\mathrm{m}}/T_1} + k_2 e^{-t_{\mathrm{m}}/T_2} + \tau \tag{24}$$

where $\tau = T_{\mathrm{R}} - T_1 - T_2$.

Solving (24) yields the over-limit time for second-level slope disturbances.

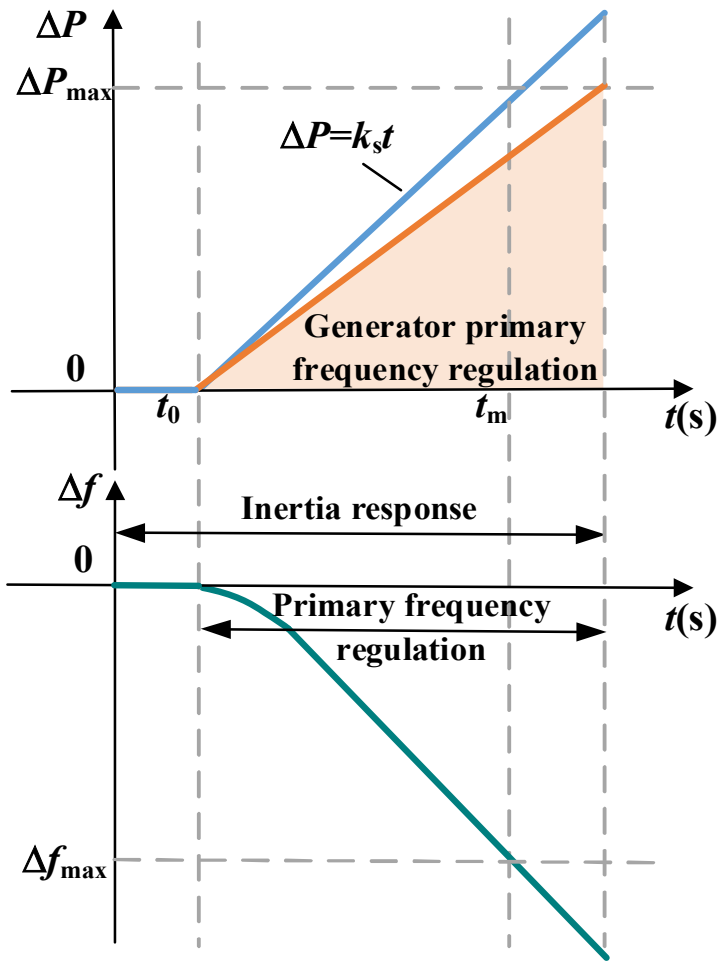

Fig. 6. Frequency response caused by a second-level slope disturbance.

## *D. Minute-Level Slope Disturbances*

For minute-level slope disturbances, the time required for the frequency deviation to reach its transient limit is much longer than in the above cases. The corresponding frequency response is shown in Fig. 7.

The frequency evolution can be divided into two stages. Before $t_1$, the primary frequency regulation of generators and loads acts cooperatively, and the system operates at a new quasi-steady-state operating point. At $t_1$, the primary frequency-regulation capacity of generators is exhausted. After $t_1$, only load frequency regulation remains effective, and the frequency deviation deteriorates progressively as the disturbance continues to grow.

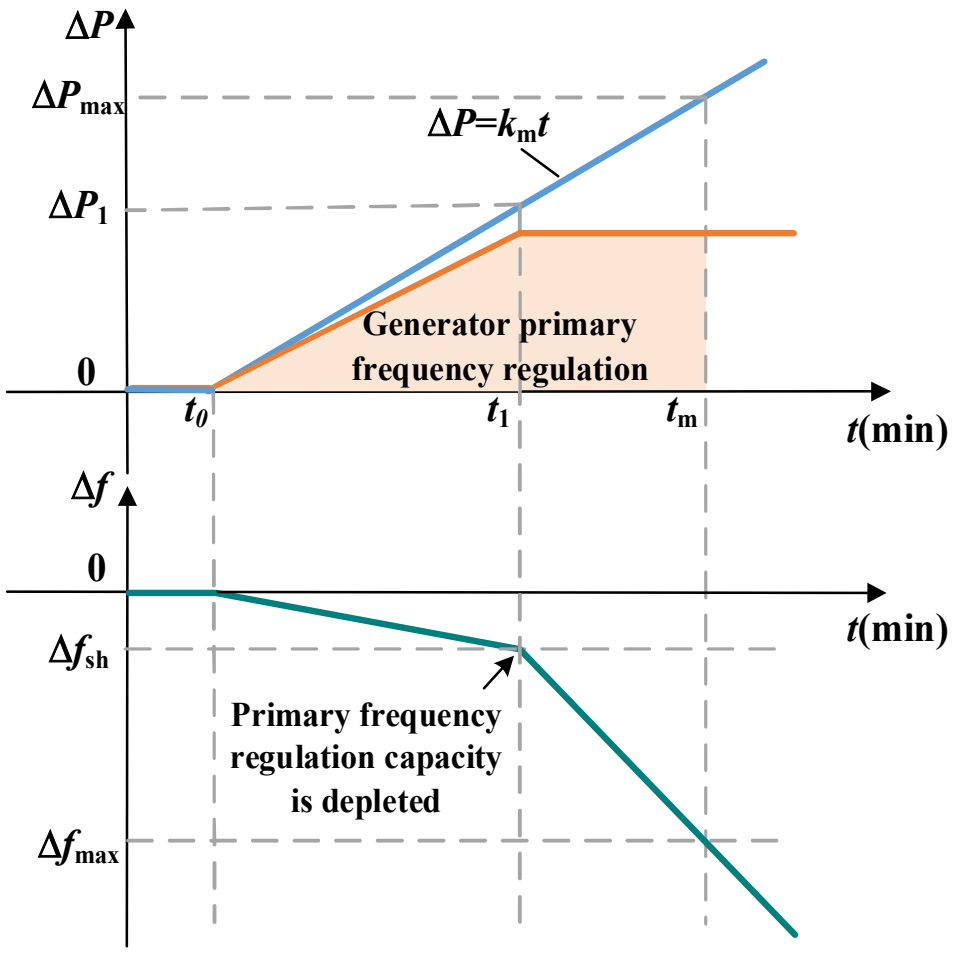

Fig. 7. Frequency response caused by a minute-level slope disturbance.

The generator rotor motion equation is:

$$\begin{cases} \dfrac{\mathrm{d}\delta}{\mathrm{d}t} = \omega - \omega_{\mathrm{N}} \\ \dfrac{\mathrm{d}\omega}{\mathrm{d}t} = \dfrac{\Delta M}{T_{\mathrm{J}}} \approx \dfrac{\Delta P_{\mathrm{m}} - \Delta P_{\mathrm{e}}}{T_{\mathrm{J}}} \end{cases} \tag{25}$$

where $T_J$ is the system inertia time constant ($T_J = 2H_{sys}$), $\delta$ is the rotor electrical angle, $\Delta P_m$ and $\Delta P_e$ are the incremental mechanical and electromagnetic power, $\omega$ is the rotor angular speed, and $\omega_N$ is the nominal angular speed.

For $t > t_1$, the generator primary frequency-regulation capability has been fully utilized:

$$\begin{cases} \Delta P_{\mathrm{m}} = 0 \\ \Delta P_{\mathrm{e}} = \Delta P + \Delta P_{\mathrm{L}} \end{cases} \tag{26}$$

After $t_1$, the additional power imbalance produced by the continuously increasing disturbance power is:

$$\Delta P = k_{\mathrm{m}}(t - t_1) \tag{27}$$

Combining (26) and (27) yields:

$$\Delta P = -(T_{\mathrm{J}}\frac{\mathrm{d}\Delta f}{\mathrm{d}t} + K_{\mathrm{Lsys}}\Delta f) \tag{28}$$

Equation (28) is a first-order linear non-homogeneous differential equation. Its general solution for the frequency deviation is:

$$\Delta f(t) = Ce^{-K_{\mathrm{Lsys}}t/T_{\mathrm{J}}} + \frac{k_{\mathrm{m}}}{K_{\mathrm{Lsys}}}(t_1 - t + \frac{T_{\mathrm{J}}}{K_{\mathrm{Lsys}}}) \tag{29}$$

At $t = t_1$, the incremental frequency deviation caused by the newly generated power imbalance is zero, which determines the integration constant $C$ as:

$$C = -e^{K_{\mathrm{Lsys}}t_1/T_{\mathrm{J}}}\frac{k_{\mathrm{m}}T_{\mathrm{J}}}{K_{\mathrm{Lsys}}^{2}} \tag{30}$$

Substituting (30) into (29), and noting that the frequency deviation at $t_1$ has reached $\Delta f_{sh}$, the post-exhaustion frequency response can be expressed as:

$$\Delta f(t) = -\frac{k_{\mathrm{m}}T_{\mathrm{J}}}{K_{\mathrm{Lsys}}^{2}}e^{K_{\mathrm{Lsys}}(t_1 - t)/T_{\mathrm{J}}} + \frac{k_{\mathrm{m}}}{K_{\mathrm{Lsys}}}(t_1 - t + \frac{T_{\mathrm{J}}}{K_{\mathrm{Lsys}}}) + \Delta f_{\mathrm{sh}} \tag{31}$$

Replacing $\Delta f(t)$ and $t$ by $\Delta f_{max}$ and $t_m$, respectively, the over-limit $t_m$ can be obtained using numerical methods. This gives the time to frequency-limit violation under minute-level slope disturbances.

### *E. Implementation Procedure of the Proposed Quantitative Assessment Method*

The implementation steps for quantitative discrimination of frequency stability under unanticipated disturbances are summarized in Fig. 8.

*1) Short-Term and Step Disturbances*

Once a short-term or step disturbance is identified, the disturbance power $\Delta P_0$ is estimated from generator and disturbance-point voltage measurements using (3)–(6). The maximum tolerable disturbance power $\Delta P_{max}$ is then calculated from the transient frequency-deviation constraint for short-term disturbances via (14), and from both steady-state and transient constraints for step disturbances via (15)–(20). Comparing $\Delta P_0$ with $\Delta P_{max}$ determines whether the system frequency remains within the prescribed limits.

*2) Second-Level Slope Disturbances*

For a second-level slope disturbance, the disturbance-power slope is first evaluated using (7). The values of $k_s$ and the transient frequency-deviation limit $\Delta f_{max}$ are then substituted into the slope-type SFR model (22)–(24) to compute the time $t_m$ at which the frequency deviation reaches its limit. This over-limit time provides a quantitative indicator of the frequency stability margin under second-level slope disturbances.

*3) Minute-Level Slope Disturbances*

For a minute-level slope disturbance, the disturbance-power slope $k_m$ at the exhaustion-related frequency threshold is obtained from the power-balance condition using (9) together with the thresholds defined in (1)–(2). Based on (31), the time $t_m$ for the frequency deviation to reach its limit is then predicted. This enables the operator to assess how long the system can withstand a slowly evolving disturbance before violating the frequency-security constraints.

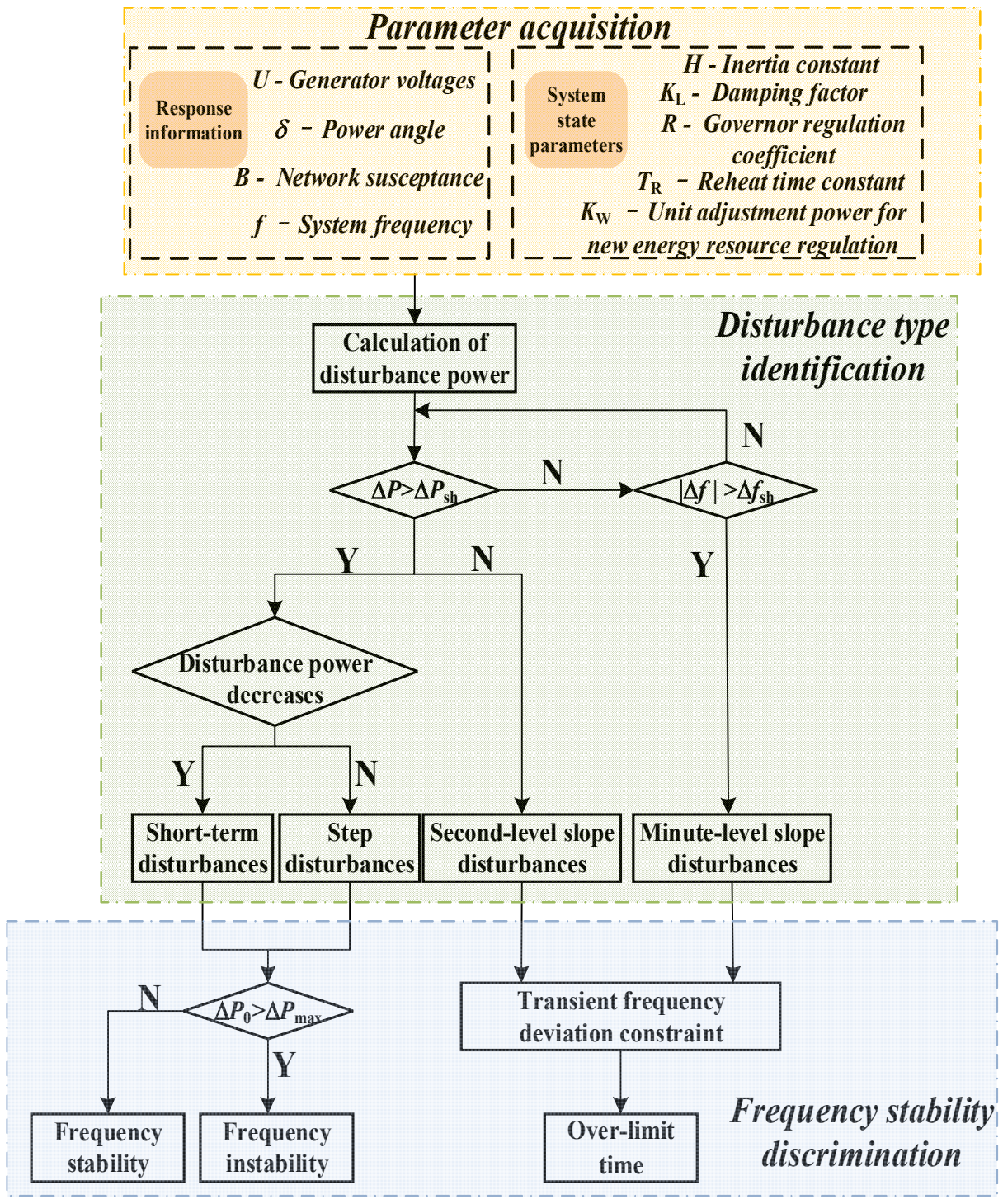

Fig. 8. Flow chart of response-based quantitative frequency stability assessment under unanticipated disturbances.

## V. Case Studies and Verification

### A. Test System Description

This section validates the proposed response-based quantitative frequency stability assessment method using the frequency-stability benchmark system developed by the China Electric Power Research Institute, denoted as CSEE-FS [27]. The benchmark represents a realistic AC–DC hybrid power system with 47 buses, 31 AC transmission lines, and three ±500 kV HVDC links. The generation portfolio consists of seven conventional units with a total installed capacity of 5400 MW and 6900 MW renewable generation from wind and photovoltaic plants. The installed renewable share reaches 56.1%. The total active load is 4852.1 MW and the three HVDC links export a total power of 4000 MW. The main 500 kV backbone network and the locations of conventional and renewable plants are shown in Fig. 9.

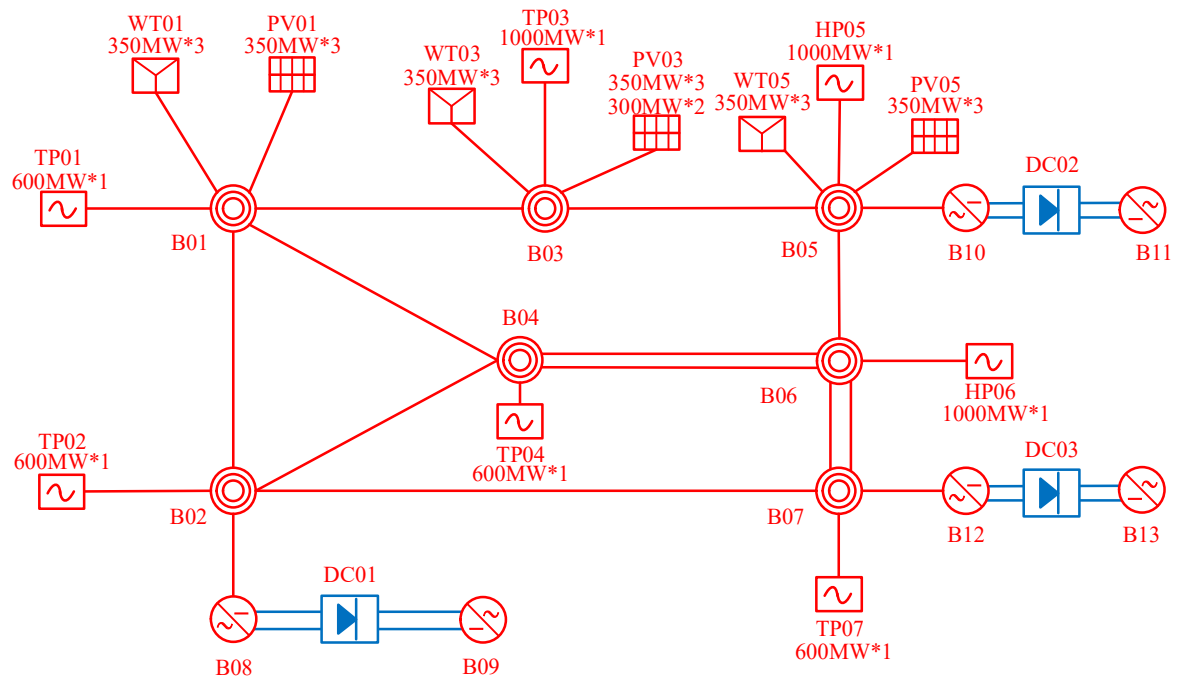

Fig. 9. System topology of CSEE-FS.

The low-frequency operating scenario is selected as the base operating point. The equivalent inertia time constant and the load-frequency damping coefficient are $T_J = 2.68$ s and $K_L = 2.0$, respectively. The aggregated SFR parameters are chosen as $K_w = 3$, $R = 0.05$, $F_H = 0.3$, and $T_R = 10$ s. The primary frequency-regulation limits of conventional generators and renewables are both set to $m = 6\%$ and $n = 6\%$ of their respective online capacities. The unbalanced-power distribution time at disturbance onset is set to $T = 1.0$ s. To investigate the impact of different renewable penetration levels, four operating conditions are further constructed by adjusting the renewable and synchronous generation capacities. The corresponding system inertia constant $H_{sys}$ and load-frequency damping coefficient $K_{Lsys}$ are listed in Table III.

TABLE III
Parameters of Operating Conditions with Different Renewable Penetration Levels

| Operating condition | Renewable power output share (%) | Renewable capacity (MW) | Synchronous capacity (MW) | $H_{sys}$ (s) | $K_{Lsys}$ (pu) |
|---|---|---|---|---|---|
| M1 | 52.28 | 7380 | 4920 | 1.340 | 3.00 |
| M2 | 60.51 | 8542 | 4072 | 1.221 | 3.47 |
| M3 | 70.45 | 9945 | 3047 | 0.916 | 4.04 |
| M4 | 80.06 | 11302 | 2056 | 0.610 | 4.59 |

For minute-level slope disturbances, the critical disturbance-power slope is set to $k_1 = 7.3$ MW/s. Accordingly, the disturbance-power threshold and the exhaustion-related frequency threshold are $\Delta P_{sh} = 7.3$ MW and $\Delta f_{sh} = 0.33$ Hz.

### B. Verification of Short-Term Disturbances

All simulations are carried out in PSD-BPA. Based on operating condition M1, three disturbance scenarios are established as summarized in Table IV, where the LVRT duration is set to 0.4 s, 0.5 s, and 0.6 s, respectively, and the active-power recovery rate of the renewable plant is $K_R = 2000$ MW/s. With the admissible maximum transient frequency deviation set to $|\Delta f_{max}| = 0.75$ Hz, the maximum tolerable disturbance power is calculated from (14) for each LVRT duration, providing the transient-limit-based boundary for short-term disturbances.

At $t$ = 1.0 s, a three-phase fault is applied to the 500 kV transmission line between buses B03 and B05. The LVRT control of PV03 is activated, which temporarily suppresses its active power and produces a short-term power deficit. The active-power trajectory of PV03-1 is shown in Fig. 10. Using (3), the estimated disturbance power $\Delta P_0 = 1350$ MW, which exceeds the early-stage threshold $\Delta P_{sh}$ within the early-stage identification window and then rapidly returns close to the pre-disturbance level after fault clearing, which is consistent with the definition of a short-term disturbance.

TABLE IV
Maximum Tolerable Disturbance Power and Maximum Transient Frequency Deviation For M1 Under Different LVRT Durations

| LVRT duration (ms) | Maximum tolerable disturbance power (MW) | Maximum transient frequency deviation (Hz) | Stability margin (%) |
|---|---|---|---|
| 400 | 1380.2 | -0.543 | 2.19 |
| 500 | 1253.8 | -0.789 | -7.67 |
| 600 | 1143.8 | -0.892 | -18.03 |

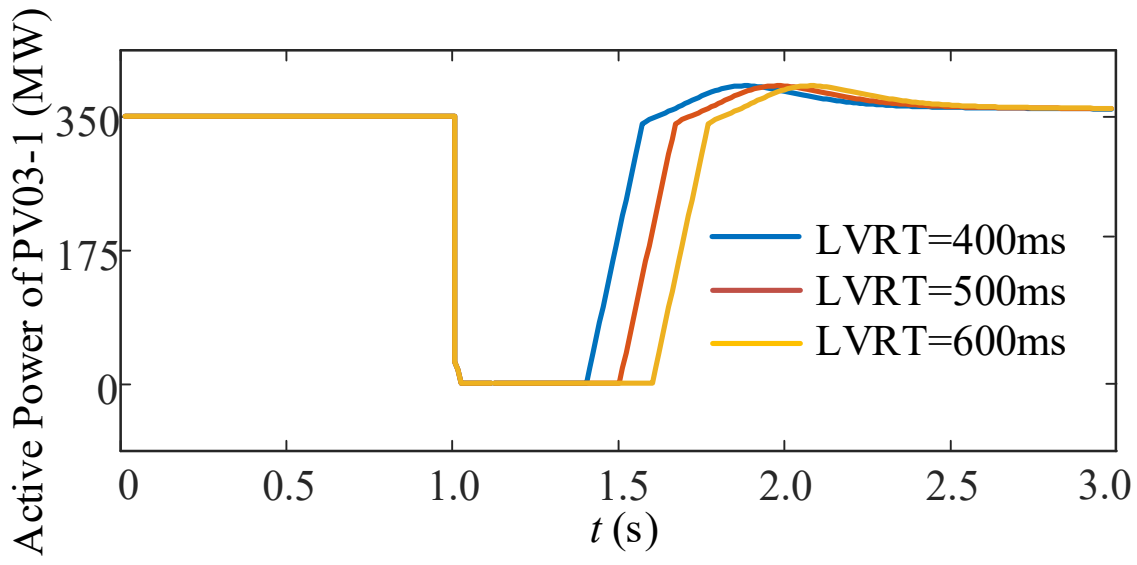

Fig. 10. Active power of PV03-1 under different LVRT durations.

During the LVRT interval, the frequency excursion is dominated by inertia because primary frequency regulation is delayed, and after clearing the steady-state deviation is negligible due to disturbance recovery. The simulated frequency trajectories in Fig. 11 show that the system remains within the transient limit only for the shortest LVRT duration, while longer LVRT durations lead to transient limit violation, which is consistent with the discrimination criterion in (14). For the three LVRT-duration cases, the maximum transient frequency deviations are 0.543 Hz, 0.789 Hz, and 0.892 Hz, respectively. Under the 1350 MW short-term disturbance, the system remains within the transient limit only when the LVRT duration is 0.4 s, while the 0.5 s and 0.6 s cases violate the transient frequency deviation constraint.

For comparison, an assessment based on the improved SFR model developed in this paper is also performed. Because the LVRT-induced deficit recovers rapidly after fault clearing, representing it in the improved SFR framework using an equivalent profile effectively treats the event as a step-type deficit of $\Delta P_0$ = 1350 MW over an overly long interval, which overestimates the accumulated deficit energy and leads to a conservative nadir prediction. The corresponding improved-SFR-predicted frequency trajectories are shown as the dashed curve in Fig. 11, which predicts limit violation for all three LVRT durations and a much deeper nadir, with the maximum transient deviation reaching 1.60 Hz. It should be noted that the deep frequency decline predicted by the step-equivalent improved SFR model is not an actual protected-system trajectory. It only indicates that the step-equivalent assessment predicts a frequency-limit violation. In practice, under-frequency protection or emergency control would be triggered once the admissible operating limit is exceeded. In contrast, the simulation results show that the 0.4 s case remains stable, indicating that preserving the finite-duration recovery feature is necessary for correct discrimination under LVRT-induced short-term disturbances.

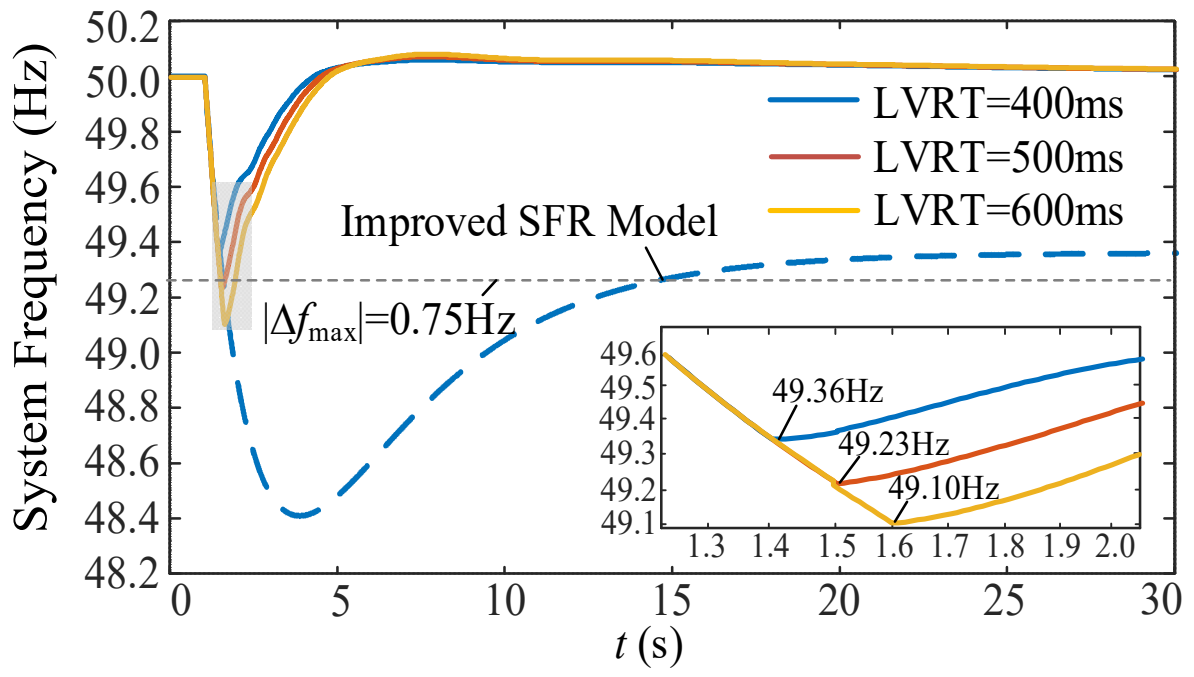

Fig. 11. System frequency responses for M1 under different LVRT durations.

Furthermore, the method is verified under operating conditions M2–M4 using the same 400-ms disturbance. The maximum tolerable short-term disturbance powers are 1292.8 MW, 1049.8 MW, and 772.2 MW, respectively. The corresponding frequency responses are shown in Fig. 12. Since the applied disturbance exceeds these limits, all three cases exhibit transient frequency-limit violations, consistent with the proposed assessment criterion.

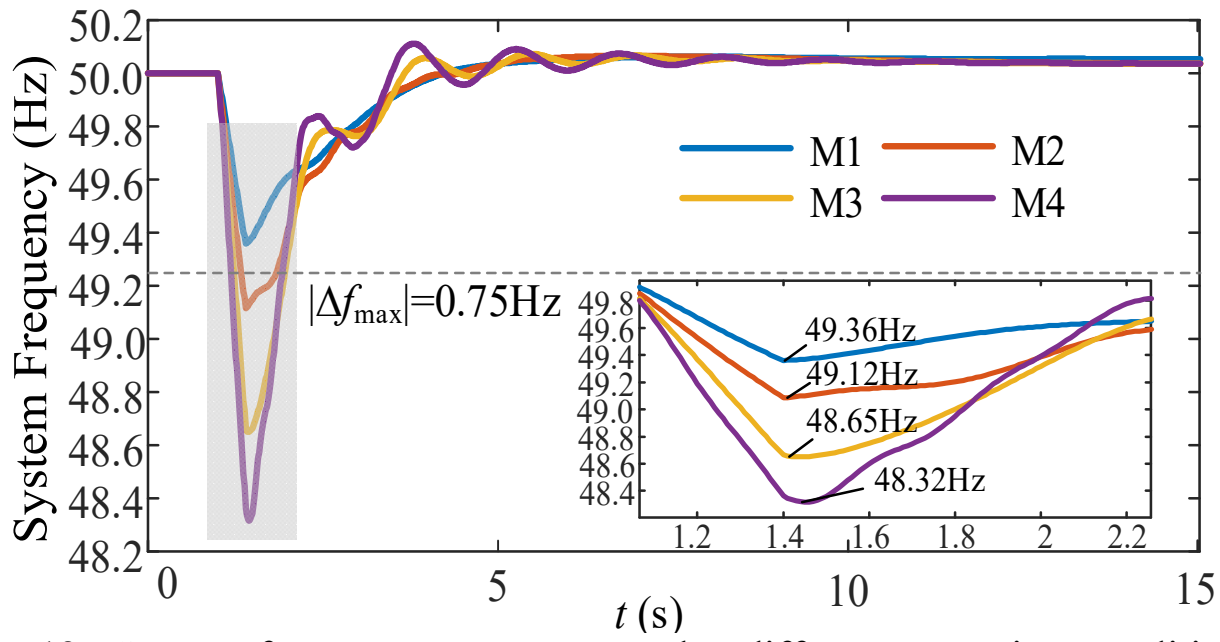

Fig. 12. System frequency responses under different operating conditions during the 400-ms LVRT disturbance.

## C. Verification of Step Disturbances

For step disturbances, the steady-state frequency-deviation limit is set to $|\Delta f_{ss}|$ = 0.2 Hz, and the maximum transient frequency-deviation limit is kept at $|\Delta f_{max}|$ = 0.75 Hz. For operating condition M1, according to (15) and (16), the critical disturbance powers under the steady-state constraint are 398.8 MW and 362.8 MW, respectively. Based on the transient nadir criterion in (17)–(19), the critical disturbance power under the transient constraint is 647.6 MW. Therefore, the overall critical disturbance power is $\Delta P_{max}$ = 362.8 MW, governed by the steady-state constraint.

To verify this result, step load-increase disturbances of 350 MW, 520 MW, and 690 MW are applied at $t$ = 1.0 s at bus B01. During the power-distribution interval, the estimated disturbance power $\Delta P$ from (6) remains approximately constant and exceeds $\Delta P_{sh}$ in all three cases, confirming step disturbances. Comparing $\Delta P$ with $\Delta P_{max}$ indicates that the 350 MW case remains within limits with a small margin, the 520 MW case violates the steady-state deviation constraint, and the 690 MW case violates both the transient nadir and steady-state constraints. The simulated frequency responses are shown in Fig. 13, and Table V reports the corresponding frequency deviations and stability margins, including 3.53% for 350 MW and −43.33% and −90.19% for 520 MW and 690 MW.

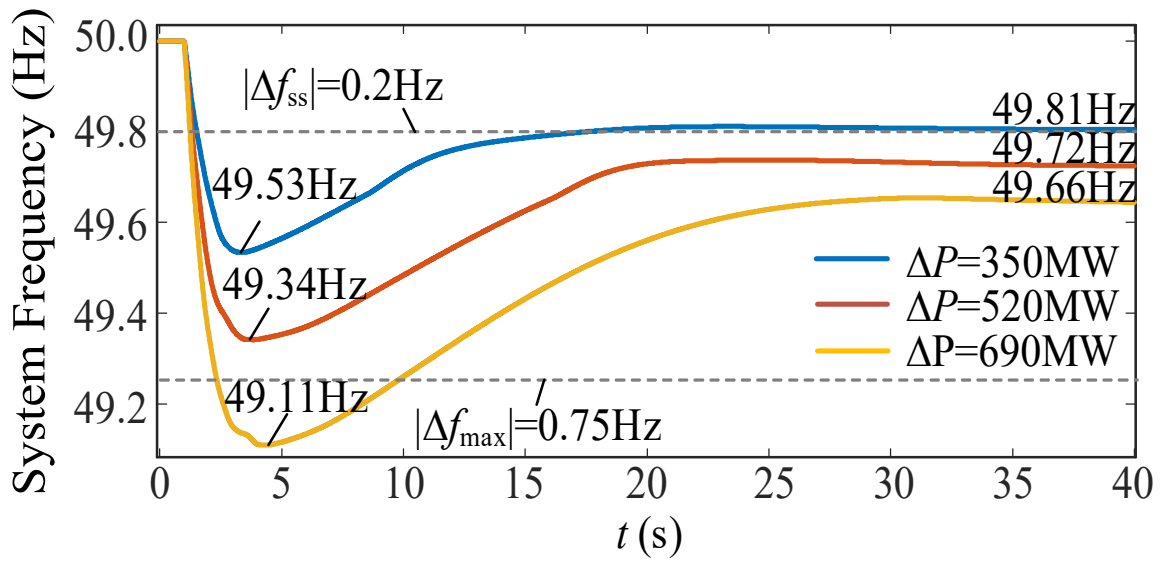

Fig. 13. System frequency responses for M1 under step disturbances with different disturbance powers.

TABLE V
FREQUENCY DEVIATIONS AND STABILITY MARGIN FOR M1 UNDER DIFFERENT STEP DISTURBANCE POWERS

| Disturbance power (MW) | Maximum transient frequency deviation (Hz) | Steady-state frequency deviation (Hz) | Stability margin (%) |
|---|---|---|---|
| 350 | -0.47 | -0.19 | 3.53 |
| 520 | -0.66 | -0.28 | -43.33 |
| 690 | -0.89 | -0.36 | -90.19 |

For comparison, the improved SFR model is also applied to the same step disturbances. As shown in Fig. 14, it yields the same accept or reject decisions as the simulation-based assessment. Quantitatively, the improved SFR model predicts the minimum and steady-state frequency deviations of −0.46 Hz and −0.19 Hz for 350 MW, −0.68 Hz and −0.30 Hz for 520 MW, and −0.89 Hz and −0.34 Hz for 690 MW, respectively, confirming that it provides correct discrimination for genuine

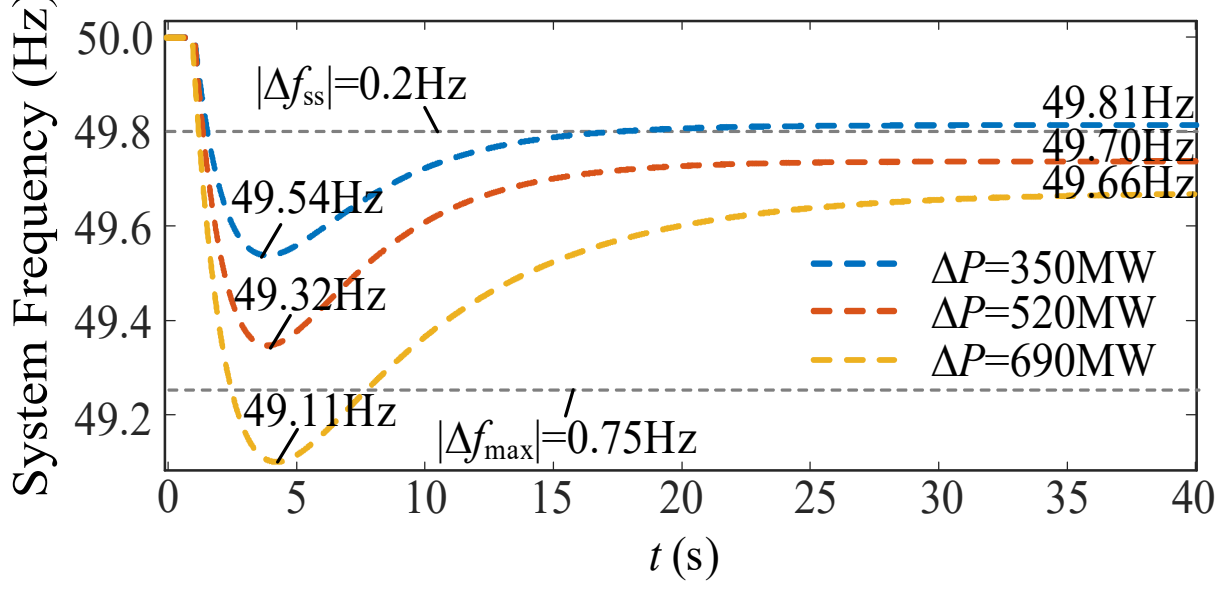

Fig. 14. System frequency responses predicted by the improved SFR model under step disturbances with different disturbance powers.

step disturbances with approximately constant power imbalance.

For operating conditions M2–M4, a 350 MW step load disturbance is applied to verify the transient frequency-deviation criterion. In all three cases, the steady-state frequency deviations exceed the prescribed limit. Therefore, the analysis focuses on whether the transient limit is violated. The critical disturbance powers under the transient frequency-deviation constraint are 570.3 MW, 444.6 MW, and 311.1 MW for M2, M3, and M4, respectively. As shown in Fig. 15, the disturbance remains below the transient critical powers for M2 and M3, and the transient frequency deviations remain within the limit, although the steady-state deviations exceed the corresponding threshold. For M4, the disturbance exceeds the transient critical power of 311.1 MW, causing both transient and steady-state limit violations. These results are consistent with the proposed quantitative assessment criterion.

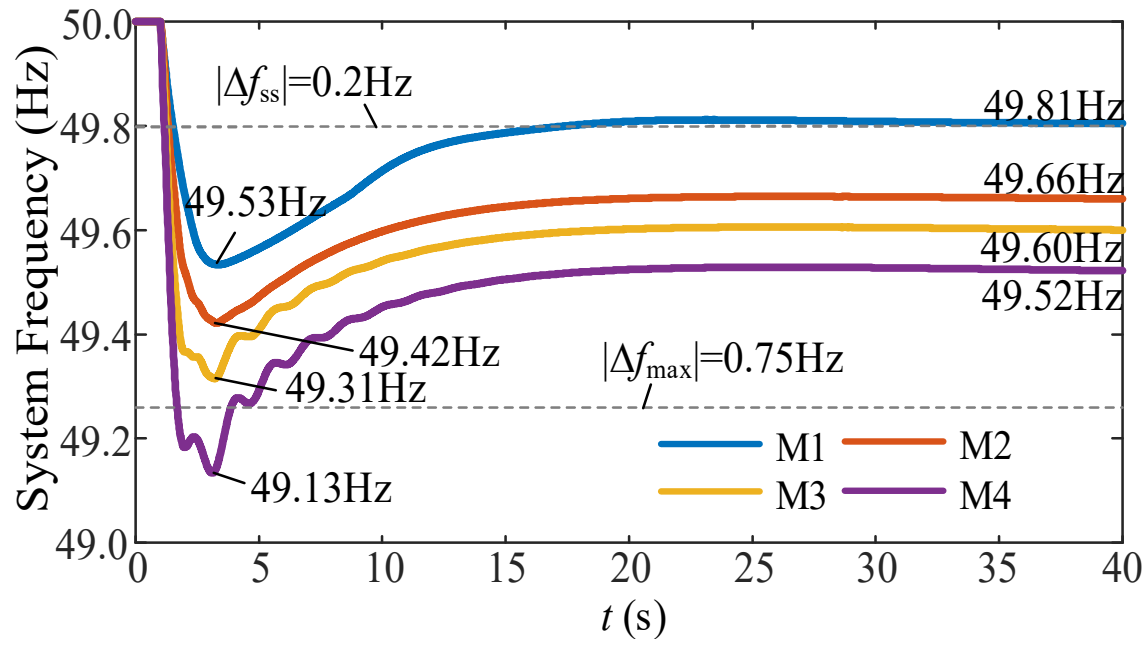

Fig. 15. System frequency responses under 350MW step disturbances for different operating conditions.

### D. Verification of Second-Level Slope Disturbances

To evaluate the influence of renewable penetration on the proposed second-level slope-disturbance assessment, the same slope-disturbance scenario is applied to the four operating conditions M1–M4. The HVDC transmission power is commanded to vary linearly from $t$ = 1.0 s, resulting in a total power change of 1200 MW over 20 s. Within the initial power-distribution interval $T$ = 1.0 s, the disturbance power $\Delta P$ computed from (6) exhibits an approximately linear growth and exceeds the threshold $\Delta P_{sh}$ at 122 ms after inception, identifying the event as a second-level slope disturbance. The disturbance-power slope from (7) is $k_s$ = 60 MW/s. Based on the identified slope disturbance, the proposed analytical model is used to predict the frequency over-limit time under M1–M4. The predicted and simulated over-limit times, together with the corresponding relative errors, are summarized in Table VI.

TABLE VI
COMPARISON BETWEEN PREDICTED AND SIMULATED OVER-LIMIT TIMES FOR THE SECOND-LEVEL SLOPE DISTURBANCE

| Operating condition | Predicted over-limit time (s) | Simulated over-limit time (s) | Relative error (%) |
|---|---|---|---|
| M1 | 13.13 | 13.65 | 3.81 |
| M2 | 10.64 | 10.91 | 2.54 |
| M3 | 8.62 | 9.31 | 8.00 |
| M4 | 6.50 | 6.96 | 7.08 |

For comparison, a step-type improved SFR criterion is applied by approximating the disturbance with an equivalent step using the effective increment within $T$. With $\Delta P$ = 60 MW as the step input, the improved SFR assessment predicts a negligible excursion and cannot provide a meaningful over-limit time, because a step model cannot capture the cumulative-growth nature of a slope disturbance. If the improved SFR model is instead driven by the same slope input, it yields essentially the same predicted trajectory as the proposed slope-based model, which is expected since the two assessments rely on the same underlying low-order frequency-response dynamics and differ only in the disturbance input form.

Fig. 16 compares the frequency trajectories obtained from the simulation, the proposed slope-based prediction, and the step-equivalent improved SFR model under the four operating conditions. The proposed method closely follows the simulated responses, whereas the step-equivalent improved SFR model deviates markedly due to its inability to represent the cumulative-growth nature of the slope disturbance. As shown in Table VI, although the prediction error slightly increases with renewable penetration, it remains within an acceptable range. These results demonstrate that the proposed method can provide effective time-to-limit prediction and support the timely implementation of corrective control measures under second-level slope disturbances.

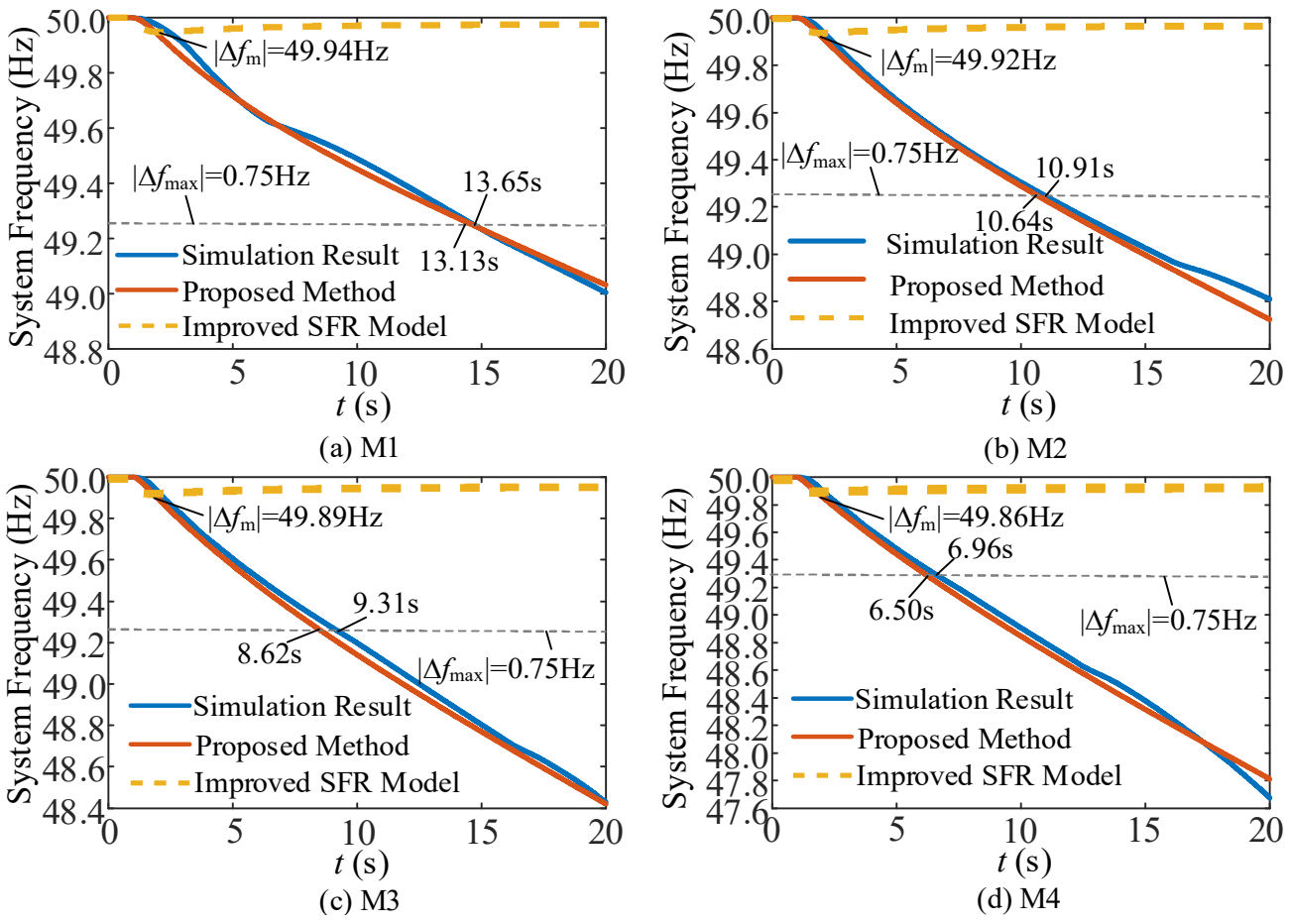

Fig. 16. System frequency responses under the second-level slope disturbance for different operating conditions.

### E. Verification of Minute-Level Slope Disturbances

To verify the proposed assessment method for minute-level slope disturbances, a continuously increasing load disturbance is applied to the four operating conditions M1–M4. The disturbance starts at $t$ = 1.0 s and the load increases linearly at a rate of 0.11% of the total active load per second. The corresponding ramp rate is approximately 5.34 MW/s, leading to a total load increase of about 1600 MW over 300 s.

Within the initial power-distribution interval, the inferred disturbance power $\Delta P$ obtained from (6) increases slowly and remains below the early-stage threshold $\Delta P_{sh}$, so the event cannot be identified as a step or a second-level slope disturbance at inception. When the frequency deviation reaches the exhaustion-related threshold, the primary frequency regulation of conventional generators is regarded as fully utilized, the event is therefore identified as a minute-level slope disturbance. The post-exhaustion assessment is then initiated from the corresponding time $t_1$ under each operating condition. The disturbance-power slope from (9) is $k_m = 5.34$ MW/s, and the proposed post-exhaustion model in (31) is used to predict the frequency over-limit time. Table VII summarizes the predicted over-limit times, simulated over-limit times, and relative errors for the four operating conditions.

TABLE VII
COMPARISON BETWEEN PREDICTED AND SIMULATED OVER-LIMIT TIMES FOR THE MINUTE-LEVEL SLOPE DISTURBANCE

| Operating condition | Predicted over-limit time (s) | Simulated over-limit time (s) | Relative error (%) |
|---|---|---|---|
| M1 | 254.2 | 261.9 | 3.03 |
| M2 | 227.5 | 238.3 | 4.75 |
| M3 | 214.7 | 220.6 | 2.75 |
| M4 | 203.1 | 211.7 | 4.23 |

Fig. 17 compares the frequency responses obtained from the simulation, the proposed minute-level slope-disturbance model, and the improved SFR model under the four operating conditions. The proposed model shows good agreement with the simulation results after the primary-regulation exhaustion instant $t_1$. It can also be observed that, as the system inertia level decreases with increasing renewable penetration, the frequency varies more rapidly both during the inertial response stage and after the exhaustion of primary frequency-regulation capability, which leads to an earlier occurrence of $t_1$ and the frequency over-limit instant $t_m$. As summarized in Table VII, the predicted over-limit times are close to the simulated results for all four operating conditions, with relative errors within an acceptable range. This confirms that the proposed method can effectively identify minute-level slope disturbances and predict the remaining time margin after the primary frequency-regulation capability is exhausted.

For comparison, the improved SFR model is also evaluated using the same minute-level slope input rather than an equivalent step input. Although this avoids the obvious bias caused by step approximation, the improved SFR model still exhibits noticeable mismatch: it predicts a later crossing of the exhaustion-related threshold, indicating that it overestimates the time required to reach $|\Delta f| = |\Delta f_{sh}|$. Moreover, its long-term frequency deviation deviates significantly from the simulation results, underestimating the continuing deterioration driven by the minute-level slope. As a result, a conventional SFR-based assessment may provide an overly optimistic remaining time margin or even miss the impending limit violation. These results demonstrate that the proposed method can provide a reliable remaining time margin for minute-level slope disturbances under different renewable penetration levels.

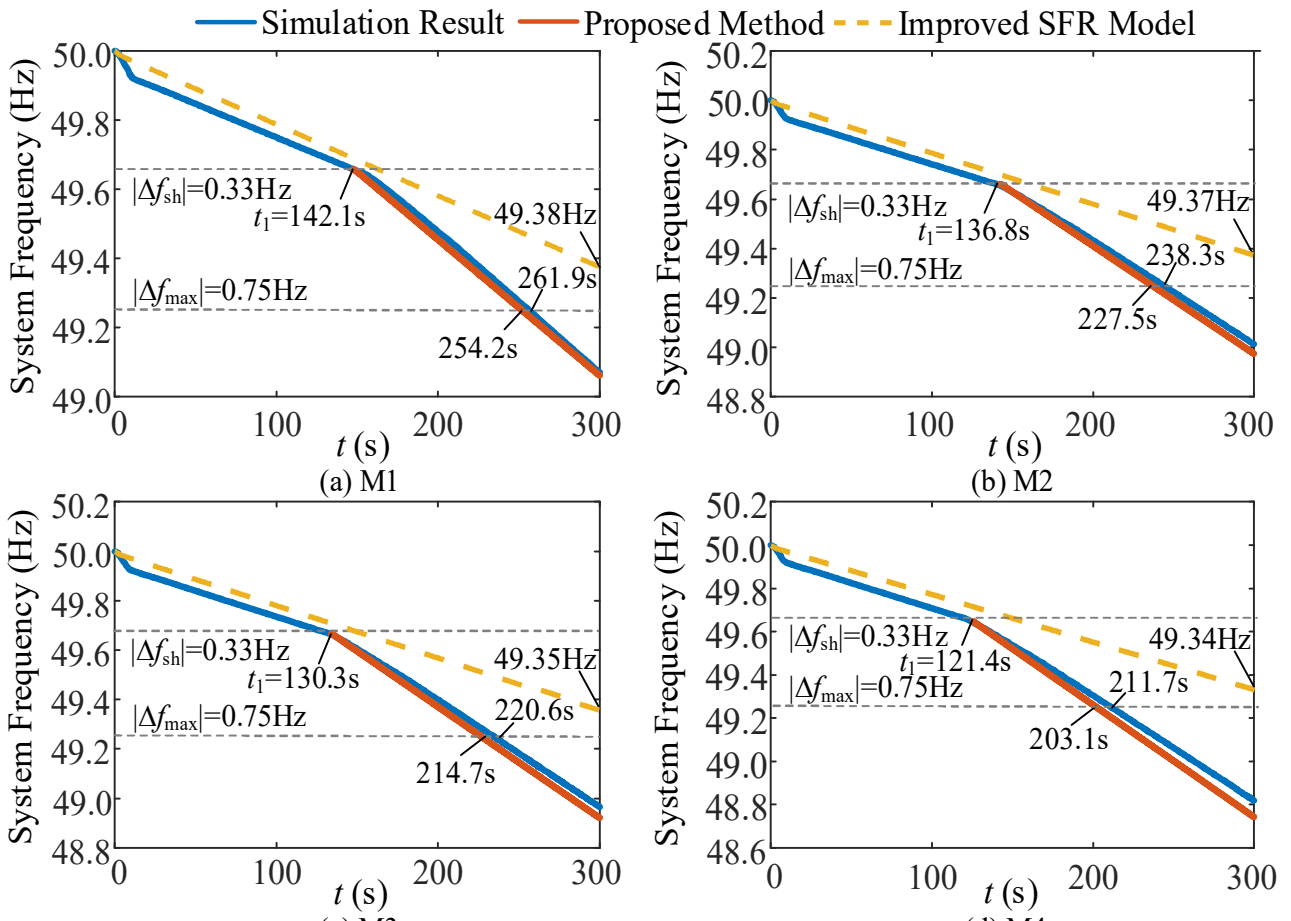

Fig. 17. System frequency responses under the minute-level slope disturbance for different operating conditions.

## F. Sensitivity and Robustness Analysis

The disturbance-power estimator in (6) depends on voltage magnitudes and phase angles. Therefore, PMU magnitude errors, phase-angle errors, and reporting-latency errors may propagate into the inferred disturbance power $\Delta P_d(t)$, and further affect the disturbance-magnitude index $\Delta P_0(t)$ and the fitted slope $k_s$. In this subsection, the influence of these errors is evaluated through a simplified error-propagation model and Monte Carlo perturbation of $\Delta P_d(t)$. The measured voltage magnitude and phase angle are modeled as:

$$\begin{aligned} \hat{E}_i(t_\ell) &= E_i(t_\ell)\left[1+\epsilon_{E_i}(t_\ell)\right] \\ \hat{\delta}_i(t_\ell) &= \delta_i(t_\ell)+\epsilon_{\delta_i}(t_\ell)-\dot{\delta}_i(t_\ell)\tau_i \end{aligned} \tag{32}$$

where $\epsilon_{E_i}(t_\ell)$ and $\epsilon_{\delta_i}(t_\ell)$ denote the voltage-magnitude error and phase-angle error, respectively. In the Monte Carlo test, they are sampled as

$$\epsilon_{E_i}(t_\ell) \sim U(-0.2\%, 0.2\%), \quad \epsilon_{\delta_i}(t_\ell) \sim U(-0.5^\circ, 0.5^\circ) \tag{33}$$

According to the first-order sensitivity of the power-transfer equation, the above magnitude and phase-angle errors are equivalently represented as a relative error in the inferred disturbance power. Considering the combined influence of voltage-magnitude errors, phase-angle errors, and aggregation effects, the equivalent power-estimation error is set as:

$$\eta_P^{(r)}(t_\ell) \sim \mathcal{N}(0, \sigma_P^2), \quad \sigma_P = 1\%, \quad |\eta_P^{(r)}(t_\ell)| \le 3\% \tag{34}$$

where $r$ denotes the Monte Carlo trial. Therefore, the perturbed disturbance-power trajectory is generated by

$$\widehat{\Delta P}_d^{(r)}(t_\ell) = \Delta P_d(t_\ell)\left[1+\eta_P^{(r)}(t_\ell)\right] \tag{35}$$

For each perturbed trajectory, the disturbance intensity and slope are recalculated as:

$$\begin{aligned} \widehat{\Delta P}_0^{(r)} &= \max_{t\in[t_0, t_0+T]} \left|\widehat{\Delta P}_d^{(r)}(t)\right| \\ \hat{k}_s^{(r)} &= \frac{\sum_{\ell=1}^{N}(t_\ell-\bar{t})\left[\widehat{\Delta P}_d^{(r)}(t_\ell)-\overline{\widehat{\Delta P}}_d^{(r)}\right]}{\sum_{\ell=1}^{N}(t_\ell-\bar{t})^2} \end{aligned} \tag{36}$$

For minute-level slope disturbances, the disturbance-intensity estimation is affected by the uncertainty of the renewable regulation limit, mainly because renewable units may reach saturation. According to (2), the error of the exhaustion-related frequency threshold caused by the uncertainty of $n$ can be expressed as:

$$\Delta(\Delta f_{sh}) = \frac{P_{NEW} f_N}{(K_G + K_W) S_N} \Delta n \tag{37}$$

where $\Delta n$ denotes the error of the renewable regulation-limit parameter. This equation shows that the uncertainty of $n$ is directly propagated to the exhaustion-related frequency threshold. Since this threshold is used to determine the exhaustion time $t_1$, the error of $n$ may further affect the accurate determination of $t_1$ and the calculation of the minute-level disturbance slope $k_m$.

To evaluate the influence of disturbance-power estimation errors, representative cases under the M1 operating condition are selected. For each case, 1000 Monte Carlo trials are performed, and the perturbed disturbance-power trajectories are generated according to (35). The short-term disturbance is selected as a representative step-like disturbance, while the second-level slope disturbance is selected as a representative slope-type disturbance. The corresponding results are shown in Figs. 18 and 19. Fig. 18 shows that the short-term disturbance still has a clear power-deficit characteristic under different LVRT durations. Fig. 19 shows the 1–2 s least-squares fitting window for $k_s$ and the Monte Carlo distribution of the fitted $k_s$ with the 68% and 95% confidence intervals.

The quantitative results are summarized in Table VIII. After considering the estimation error, the identified disturbance intensities remain within narrow intervals around their nominal values, and the disturbance-type identification results are unchanged for the short-term disturbance, step disturbance, and second-level slope disturbance. The corresponding stability margins and over-limit time are also recalculated using the perturbed results. As shown in Table VIII, the assessment results remain close to those obtained without errors, indicating that the proposed method is robust against the considered disturbance-power estimation errors.

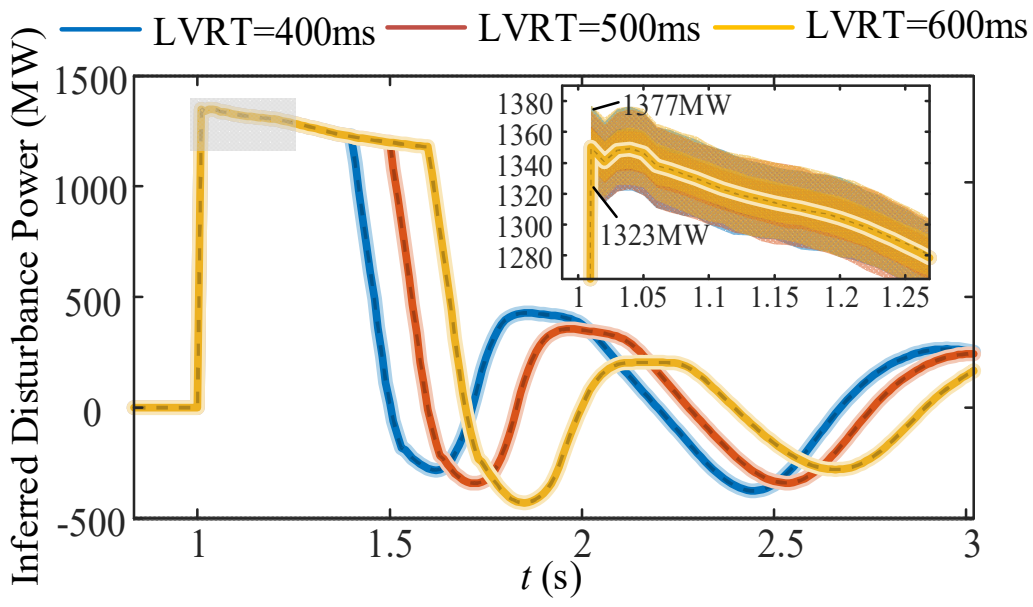

Fig. 18. Disturbance-power trajectories of short-term disturbances under different LVRT durations with power-estimation errors.

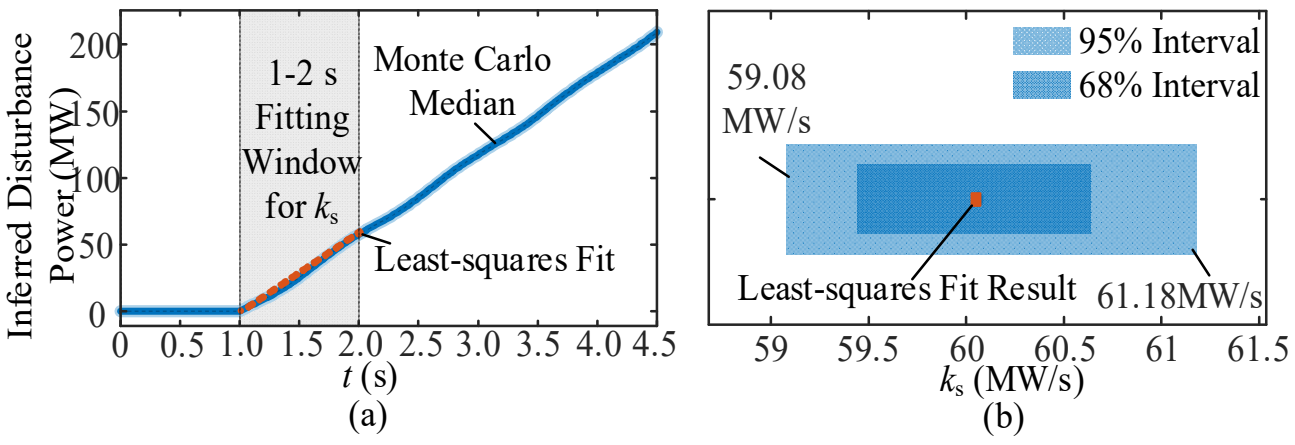

Fig. 19. Slope identification of the second-level slope disturbance under power-estimation errors.

The initial disturbance-identification results indicate that the considered power-estimation errors do not cause misclassification among different disturbance types. Therefore, for minute-level slope disturbances, the assessment accuracy is mainly limited by the uncertainty of the available frequency-regulation capacity. In practical operation, part of the renewable units may reach saturation, and the actual available regulation capacity can be lower than the preset value. If this saturation level is not known online, the preset regulation capacity may still be used to calculate $k_m$ and determine $t_1$, which leads to an overestimated disturbance slope and a conservative prediction of the over-limit time.

TABLE VIII
DISTURBANCE IDENTIFICATION AND STABILITY-ASSESSMENT RESULTS FOR THE M1 OPERATING CONDITION UNDER POWER-ESTIMATION ERRORS

| Index | Short-term disturbance | Step disturbance | Second-level slope |
|---|---|---|---|
| Actual intensity | 1350 MW | 350 MW | 60 MW/s |
| Intensity with error | [1323,1377] MW | [344,358] MW | [59.08,61.2] MW/s |
| Disturbance-type identification result | 400ms short-term disturbance | Step disturbance | Second-level slope |
| Stability margin without error | 2.19% | 3.53% | / |
| Stability margin with error | [0.23, 4.14]% | [1.32, 5.18]% | / |
| Over-limit time without error | / | / | 13.13 s |
| Over-limit time with error | / | / | 13.17 s |

To verify this effect, the M4 operating condition with the highest renewable penetration is selected as the representative case. The renewable regulation ratio $n$ is reduced from 0.06 to 0.05, 0.04, and 0.03, corresponding to regulation capacities of 801.48 MW, 688.46 MW, 574.44 MW, and 462.42 MW. The corresponding actual exhaustion thresholds are 0.330 Hz, 0.288 Hz, 0.246 Hz, and 0.204 Hz, respectively. However, the online assessment still uses the nominal threshold of 0.330 Hz to determine $t_1$. The simulated and predicted frequency responses under different renewable-regulation saturation levels are shown in Fig. 20. The results in Table IX show that the predicted over-limit time is earlier than the simulated value, indicating that the assessment becomes conservative under renewable-regulation saturation. As the available renewable regulation capacity decreases, the conservativeness becomes more obvious.

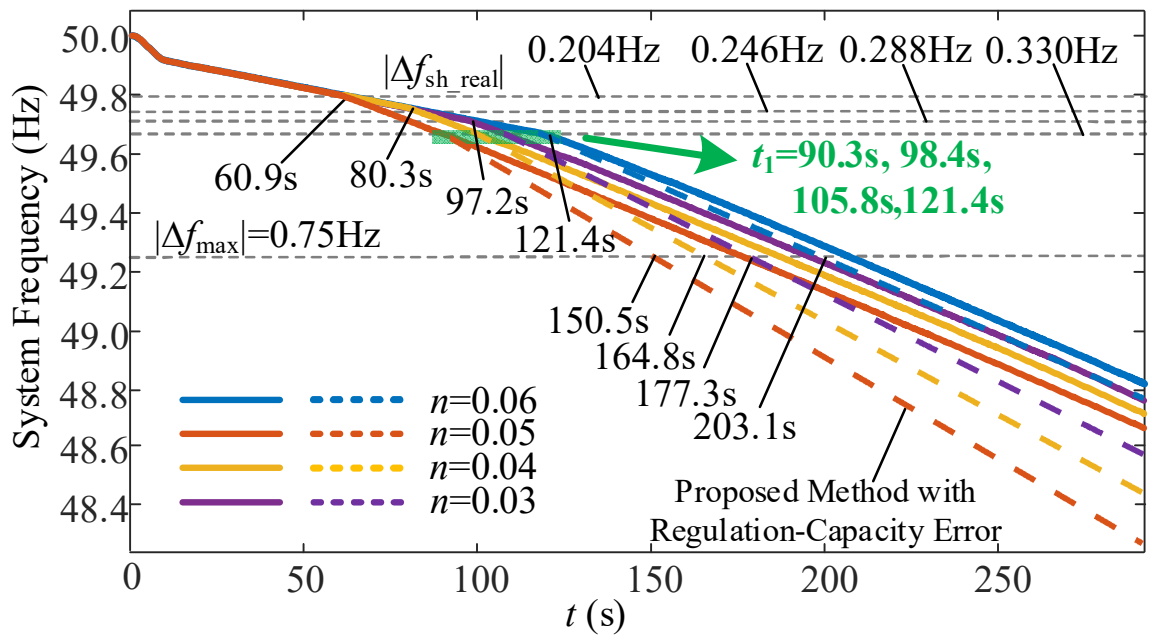

Fig. 20. Frequency responses and prediction results under renewable-regulation saturation in the M4 operating condition.

Overall, the sensitivity results show that both measurement-related disturbance-power errors and renewable-regulation-capacity uncertainty have limited influence on the disturbance-type identification results, and the proposed method can still maintain correct classification under the tested conditions. Although these uncertainties may introduce deviations in the estimated stability margin or over-limit time, the prediction results are generally conservative, which is beneficial for issuing early warnings and guiding preventive control actions.

TABLE IX
SENSITIVITY RESULTS OF MINUTE-LEVEL DISTURBANCE ASSESSMENT UNDER RENEWABLE-REGULATION SATURATION

| Renewable regulation ratio *n* | $k_m$ (MW/s) | Actual $t_1$ (s) | Detected $t_1$ (s) | Predicted over-limit time (s) | Simulated over-limit time (s) | Relative error (%) |
|---|---|---|---|---|---|---|
| 0.06 | 5.34 | 121.4 | 121.4 | 203.1 | 211.7 | 4.23 |
| 0.05 | 6.12 | 97.2 | 105.8 | 177.3 | 195.8 | 10.43 |
| 0.04 | 6.59 | 80.3 | 98.4 | 164.8 | 187.2 | 13.59 |
| 0.03 | 7.18 | 60.9 | 90.3 | 150.5 | 176.8 | 17.48 |

## VI. CONCLUSION

This paper develops a response-based quantitative frequency-stability assessment framework for high-renewable power systems under multi-scale unanticipated disturbances. By inferring disturbance power from measured generator electrical responses, the proposed method identifies disturbance types and evaluates class-dependent stability margins within a unified analytical framework.

1) A response-based disturbance classification scheme is established for short-term, step, second-level slope, and minute-level slope disturbances. The results show that the dominant feature of a disturbance is not only its magnitude, but also its duration and rate of change. By jointly using the early-stage power threshold $\Delta P_{sh}$ and the exhaustion-related frequency threshold $\Delta f_{sh}$, the proposed method can distinguish fast transient disturbances from slowly evolving minute-level disturbances that cannot be identified from early-stage power increments alone.

2) Analytical frequency-stability criteria are derived for different disturbance classes. For short-term disturbances, the proposed energy-accumulation criterion preserves the finite-duration recovery feature and avoids the excessive conservativeness caused by treating the event as a permanent step disturbance. For step disturbances, the derived critical power boundary provides explicit stability margins under both steady-state and transient frequency-deviation constraints. For second-level and minute-level slope disturbances, the proposed models provide direct time-to-limit prediction, which is more suitable than step-equivalent assessment for continuously increasing power deficits. The derivation method in this paper can be further extended to complex disturbances by applying the energy-accumulation criterion to short-duration stages, the improved SFR-based model to stages governed by primary frequency regulation, and the post-exhaustion model to slowly evolving stages after regulation saturation.

3) Case studies on the CSEE-FS benchmark system verify the effectiveness of the proposed method. The results show that the proposed criteria provide correct stability discrimination for LVRT-induced short-term disturbances, step-load disturbances, and slope disturbances under different renewable penetration levels. In particular, for minute-level slope disturbances, the post-exhaustion rotor-motion model captures the continued frequency deterioration after primary frequency-regulation capability is exhausted, whereas the conventional SFR-based assessment may overestimate the remaining time margin.

4) The main error sources are analyzed through sensitivity and robustness tests, including disturbance-power estimation errors caused by voltage magnitude, phase-angle, and reporting-latency errors, as well as renewable-regulation-capacity uncertainty caused by saturation. The results show that these errors do not change the disturbance-type identification results under the tested conditions. Although they may affect the estimated stability margin and over-limit time, the prediction is generally conservative, which is beneficial for early warning and preventive control.

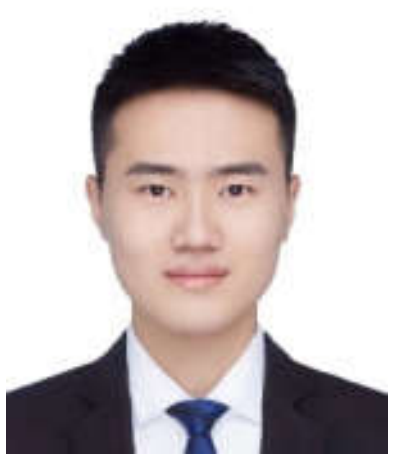

**Jinhui Chen** received the B.Eng. degree in electrical engineering from North China Electric Power University, Beijing, China, in 2021. He is currently pursuing the Ph.D. degree with the Power System Department, China Electric Power Research Institute, Beijing, China. During 2025–2026, he was a Visiting Ph.D. Student with the School of Electrical and Electronic Engineering, University College Dublin, Dublin, Ireland. His research interests include power system stability analysis and control.

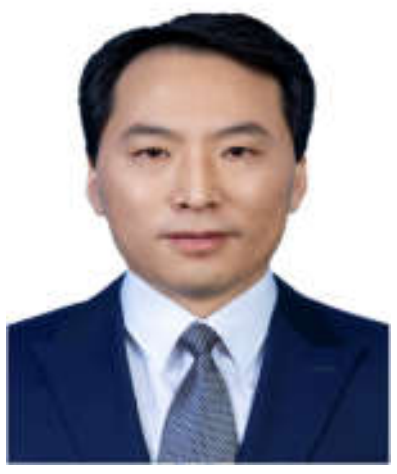

**Huadong Sun** received B.Eng. and M.S. degrees in Electrical Engineering from Shandong University, Jinan, China, in 1999 and 2002, respectively. He received the Ph.D. degree in Electrical Engineering from China Electric Power Research Institute, Beijing, China in 2005. His research interests include power system analysis and control, stability of power-electronic-based power system, and large-scale renewable power generation and integration.

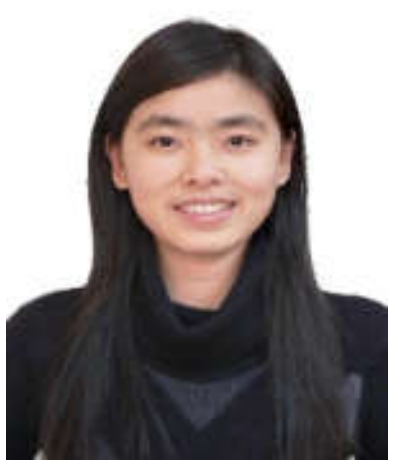

**Ping Wu** received the B.Eng. and M.S. degree in electrical engineering from Wuhan University, Wuhan, China, in 2008 and 2010, respectively. She is currently an excellent technical expert of CEPRI. Her research interests include power system analysis and control.

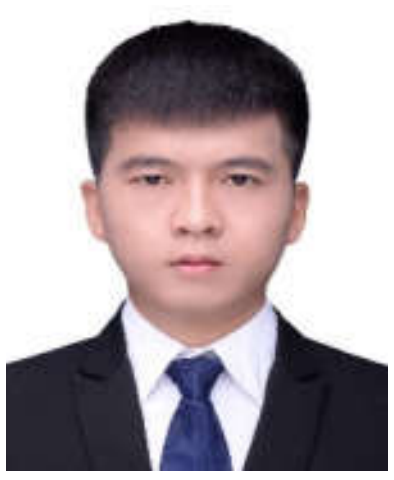

**Baocai Wang** received the B.Eng. and M.S. degrees in electrical engineering from Northeast Electric Power University, Jilin, China, in 2015 and 2018, respectively. He received the Ph.D. degree from China Electric Power Research Institute (CEPRI), Beijing, China, in 2022, in electrical engineering. His research interests include power system inertia estimation and power system frequency stability.

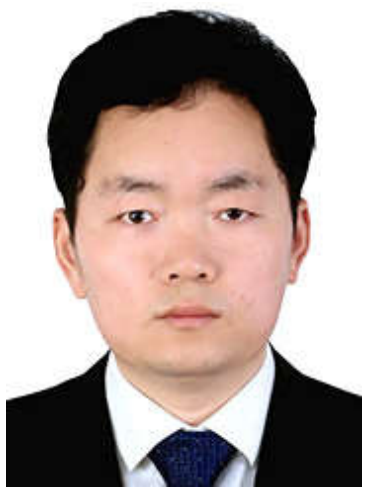

**Bing Zhao** received B.Eng. and M.S. degrees in Electrical Engineering from Xi'an University of Technology, Xi'an, China, in 2003 and 2006, respectively. He received the Ph.D. degree from China Electric Power Research Institute, Beijing, China, in 2009. Currently, he is a Senior Engineer at China Electric Power Research Institute, Beijing, China. His research interests include power system stability analysis and control, load modelling, renewable power generation and integration, and operation and control of the DC grid.